\documentclass[preprintnumbers,amsmath,amssymb]{revtex4}
\usepackage{graphicx}
\usepackage{bm}

\newcommand{\be}{\begin{equation}}
\newcommand{\ee}{\end{equation}}
\newcommand{\lra}[1]{\langle #1 \rangle }
\begin{document}
\title{Mesoscopic lattice Boltzmann modeling of soft-glassy systems: theory and simulations}
\author{R. Benzi $^{1}$, M. Sbragaglia$^{1}$, S. Succi$^{2}$, M. Bernaschi$^{2}$, S. Chibbaro$^{3}$ \\
$^{1}$ Department of Physics and  INFN, University of ``Tor Vergata'', Via della Ricerca Scientifica 1, 00133 Rome, Italy\\  
$^{2}$ Istituto per le Applicazioni del Calcolo CNR, Viale del Policlinico 137, 00161 Roma, Italy \\
$^{3}$ Department of Physics, University of `'Roma 3'', Via della Vasca Navale, 00146 Rome, Italy}

\begin{abstract} 
A multi-component lattice Boltzmann model recently introduced (R. Benzi {\it et al.}, {\it Phys. Rev. Lett.} {\bf 102}, 026002 (2009)) to describe some dynamical  behaviors of soft-flowing materials is theoretically analyzed.  Equilibrium and transport properties are derived within the framework of a continuum free-energy formulation, and checked against numerical simulations. Due to the competition between short-range inter-species repulsion and mid-range intra-species attraction, the model is shown to give rise  to a  very rich configurational   dynamics  of the density   field,  exhibiting  numerous  features  of   soft-flowing  materials, such as long-time relaxation due to caging effects,  enhanced viscosity and structural arrest, ageing under moderate shear  and shear-thinning flow above a critical shear threshold.
\end{abstract}

\maketitle

\section{Introduction}
The study of the rheology of flowing soft systems, such as emulsions, foams, gels, slurries, colloidal glasses and related complex fluids, is gaining an increasing role in modern science and engineering \cite{Larson99,Coussot05,Chaikin95,Lyklema91,Evans99,Degennes79,Doi86,Grosberg94}.  From the theoretical standpoint, much of the fascination of these systems stems from the fact that they do not fall within any of three basic states of matter, gas-liquid-solid, but live rather on a moving border between them. Foams are typically a mixture of gas and liquids, whose properties can change dramatically with the changing proportion of the two; wet-foams can flow almost like a liquid, whereas dry-foams may conform to regular patterns, exhibiting solid-like behavior \cite{Weaire99}. Emulsions can be paralleled to bi-liquid foams, with the minority species dispersed  in the dominant (continuous) one.      The behavior and, to same extent, the very existence itself of both foams and emulsions  are vitally dependent on surface tension, namely the interactions that control the physics at the interface between different phases/components. Indeed, the presence of surfactants, i.e. a third constituent with the capability of lowering surface tension, has a profound impact on the behavior of foams and emulsions; by  lowering the surface tension, surfactants can greatly facilitate mixing, a much sought-for property in countless practical endeavors, from oil-recovery, to chemical and biological applications.   Another basic property of foams and emulsions is {\it metastability/disorder}. Indeed, in most instances, these materials consist of a disordered collection of droplets/bubbles with a broad distribution of sizes, randomly mixed and arranged, which do not correspond to the (global) minimum of any thermodynamic function. This is even truer in the case of complex {\it flowing} systems, which  live consistently out of (thermodynamic) equilibrium. As a result, they exhibit a number of distinctive features, such as long-time relaxation, anomalous viscosity, aging behavior, whose quantitative description is calling for profound extensions of non-equilibrium statistical mechanics  \cite{Russel89,Poole92,Sollich97,Larson99,Eckert02,Sciortino02,Pham04,Guo07,Schall07,Lu08}. The study of these phenomena sets a pressing challenge for computer simulation as well, since characteristic time-lengths of disordered fluids can escalate tens of decades over the molecular time scales \cite{Allen90,Frankel96,Binder97}. In addition, tracking the time evolution of complex interfaces represents a serious hurdle for traditional discretization techniques. These split into two broad categories: Eulerian and Lagrangian. In Eulerian methods, the physical observables are attached to a fixed grid and monitored as they change in time at each grid location. Lagrangian methods, on the contrary, "go with the flow", i.e. the degrees of freedom are attached to the moving fields, and most notably to the critical regions of the flow where the most abrupt changes take place (interfaces). As usual, both methods have their merits and pitfalls. Lagrangian methods do not waste degrees of freedom on uninteresting regions of the flow;  however, since the grid adapts to the changing fields, when these changes are too abrupt the numerics is forced to ad-hoc readjustments (grid-rezoning) which may eventually fail and lead to collapse of the numerics \cite{ALE74} . Eulerian methods are free from these problems, because the interface is not tracked, but just tagged as the region where strong gradients are detected. The downside is that very high resolution is needed around the interface, for otherwise excessive smoothing (numerical diffusion) results (diffuse-interface) \cite{VOF99} . A special variant of Eulerian methods does not attempt to resolve the interface, which is treated as a zero-thickness mathematical interface, across which jumps of the observables are specified. A proper handling of the discontinuities, and the avoidance of spurious oscillations, is however a non-trivial task \cite{Quarteroni07}.

On a more microscopic scale, often too small for hydrodynamic purposes, to date
the most credited techniques for complex flowing materials are Molecular Dynamics and Monte Carlo simulations \cite{Allen90,Frankel96,Binder97}.   Molecular Dynamics in principle provides a fully ab-initio description of the system, but  it is limited to space-time scales significantly shorter than experimental ones.  Monte Carlo methods are somehow less affected by this limitation, since they can be designed in compliance with accelerated-dynamic sampling rules. 
However, these rules meet with some difficulties in accounting for hydrodynamic interactions \cite{Kob02}.   As a result, neither MD nor MC can easily take into account the non-equilibrium dynamics of complex flowing materials, such as micro-emulsions, on space-time scales of hydrodynamic interest.   Besides these general techniques, a number of specialized methods are also available, such as dissipative particle dynamics \cite{Hoogerbrugg92}  and others.  More in details,  a special kind of  Molecular Dynamics (MD) for dry  granular, Stokesian Dynamics (SD)   for    viscous   suspensions   and    Bubble-Model   for   foams \cite{Cun79,Dur87,Duri95}  have  been   developed  with  several  adjustable particle  interactions  in  order   to  have  a  good  agreement  with experiments \cite{GDR04,Bra01,Hol05}. By  contrast, the combination of both particle  deformation and viscous  flow has  not been  fully described yet, although it is central  in such materials as foams and emulsions. For  macroscopic  complex   flows,  particularly  interesting  is  the approach  by Doi {\it  et al.}  \cite{DOI}, which  construct a  set of evolution equations for the  volume fraction of the oriented interface elements within a  complex flows, and more recently  the soft dynamics approach \cite{Gay08}.

In the last decade, a new class of mesoscopic methods, based on minimal lattice formulations of Boltzmann's kinetic equation, have captured significant interest as an efficient alternative to continuum methods based on the discretization of the Navier-Stokes equations for non-ideal fluids \cite{Mcnamara98,Higuera89a,Higuera89b,Benzi92,Chen98,Gladrow00}.  A very popular mesoscopic technique is the   pseudo-potential-Lattice-Boltzmann (LB) method, developed over a decade  ago by Shan \& Chen  \cite{SC_93,SC_94}. In the SC method, potential energy interactions are represented through a density-dependent mean-field pseudo-potential, $\Psi(\rho)$, and  phase separation is achieved by imposing a short-range attraction between the light and dense phases.  In this work, we discuss extensions of two-species, mesoscopic lattice Boltzmann model which prove capable of reproducing some features of flowing soft- materials, such as  structural arrest, anomalous viscosity, cage-effects and ageing under shear \cite{Benzi09}.  The key feature of the model is the capability to investigate the rheology of these systems  on space-time scales of hydrodynamic interest at an affordable computational cost. Among others, this model shows the first evidence of mesoscopic cage formation and rupture within a hydrodynamic lattice Boltzmann description.

The present work is organized in two major parts: Theory and Numerical Results.
In section II, we provide the basic elements of the multicomponent lattice kinetic model with multi-range non-ideal interactions, short-range attraction and mid-range repulsion.  In section III, we derive the macroscopic equations associated with the large-scale hydrodynamic limit of the kinetic model.  In section IV and V, we present an explicit calculation of the equilibrium (equation of state) and transport (surface-tension) properties, both for the case of intra-species repulsion alone, as well as its combination with intra-species attraction.  In the process, we detail how the combination of this short/mid-range attractive/repulsive interactions allows to bring the surface tension down to vanishingly small values, a property which is key to the complex and heterogeneous dynamics displayed by the model, and notably by the density field.  The numerical part follows in section VI. In section VII we discuss the morphological features of the density configurations, and demonstrate the existence of long-lived metastable states resulting from the interplay/competition between short-range attraction and mid-range repulsion. In section VIII, we investigate the dynamic response of system under an external shear drive, and provide several evidences of complex behaviors, such as cage formation and rupture under shear, ageing and its disappearance above a critical shear threshold, long-term non-Newtonian shear-strain correlations and Barkhausen intermittency, namely a power-law distribution of the waiting times between sliding events events.   
In section IX we discuss the issues of sensitivity to initial conditions and finite-size effects. 
In section X, we conclude with an outlook and future perspectives for the application 
of the present model, and generalizations thereof, to a broad class of complex soft-flowing 
systems, such as foams, emulsions and similar. 
Finally, in the Appendix we provide the conversion rules from/to lattice to physical units.

\begin{figure}
\vspace{0.5cm}
\includegraphics[scale=0.5]{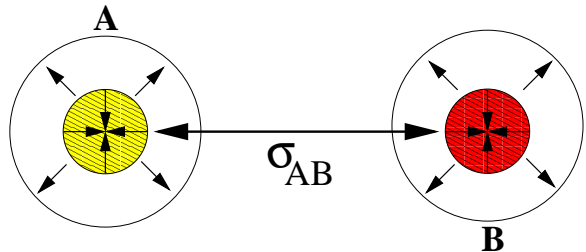}
\caption{Sketch of the interaction forces between two species, say A and B. The two components A and B interact via a repulsive pseudo-potential, which supports a surface tension $\sigma_{AB}$.  Moreover, each component experiences an attractive interaction in the first Brillouin zone and a repulsive one  acting on both Brillouin zones (see also figure \ref{fig:2} for the technical details). Each of these interactions can be tuned through a separate coupling constant.}
\label{fig:1}
\end{figure}

\section{The multi-component kinetic model}

Kinetic theory and its discrete (lattice Boltzmann) counterparts  for multicomponent fluids and  gas mixtures have received much attention in the literature \cite{GrossJackson59,Sirovich62,Hamel65,Hamel66,Sirovich66,ZieringSheinblatt66,GoldmanSirovich67,LuoGirimaji02,LuoGirimaji03,SC_93,SC_94}. Many of the kinetic models for mixtures are based on the linearized Boltzmann equations, especially the single-relaxation-time model due to Bhatnagar, Gross, and Krook -the celebrated BGK model \cite{BGK54}.  Here we shall consider the multicomponent model introduced by Shan \& Chen \cite{SC_93,SC_94}: after a brief summary the main properties of the model we will proceed to analyze the equilibrium states relevant on the hydrodynamic  scales. We start from a kinetic lattice Boltzmann equation \cite{Benzi92,Gladrow00,Chen98}  for a multicomponent fluid with $N_s$ species \cite{SD_95,SD_96} whose evolution equations over a  characteristic time lapse $\Delta t$ read as follows
\begin{equation}
f_{is}(\vec{r} + \vec{c}_i \Delta t, t + \Delta t)-f_{is}(\vec{r} , t ) =  -\frac{\Delta t}{\tau_s}[f_{is}(\vec{r} , t )-f_{is}^{(eq)}(\rho_s ,  \vec{u}+\tau_s \vec{F}_{s} /\rho_s)]   
\label{eq:be}
\end{equation}
where $f_{is}(\vec{r},t)$ is the probability density function of finding a particle of species $s=1...N_s$ at site $\vec{r}$ and time $t$, moving along the $i$-th  lattice direction defined by the discrete speeds $\vec{c}_i$ with $i=0...b$. For simplicity, the characteristic time lapse $\Delta t$ is assumed to be equal to unity in the following.  The left hand-side of (\ref{eq:be}) stands for molecular free-streaming, whereas the right-hand side  represents the time relaxation (due to collisions) towards local Maxwellian equilibrium $f_{is}^{(eq)}(\rho_s , \vec{u} )$ on a time scale $\tau_s$ \cite{Benzi92,Chen98,Gladrow00,BGK54}. The local Maxwellian is truncated at second order, an approximation that is sufficient to recover correct hydrodynamic balance in the isothermal regime
$$
f_{is}^{(eq)}(\rho_s ,  \vec{u} )=w_i^{(eq)} \rho_s \left(1+\frac{({u}_a c_{ia})}{c_S^2}+ \frac{({c}_{ia} {c}_{ib}-c_S^2 {\delta}_{ab})}{2 c_S^4}{u}_a {u}_b \right)
$$
with $c_S^2$ the square of the sound speed velocity in the model and ${\delta}_{ab}$ the Kronecker  delta with $a,b$ indicating the Cartesian components (repeated indices are summed upon).  The $w_i^{(eq)}$'s are equilibrium weights used to enforce isotropy of the hydrodynamic equations \cite{Benzi92,Gladrow00,Chen98}.  To be noted that the equilibrium for the $s$ species is a function of the local species density 
$$\rho_s(\vec{r}, t)=\sum_i f_{is}(\vec{r}, t)$$ 
and the common velocity defined as 
$$\vec{u} (\vec{r}, t)= \frac{\sum_s \frac{1}{\tau_s}  \sum_i f_{is}(\vec{r}, t) \vec{c}_i }{ \sum_s \frac{1}{\tau_s} \rho_s (\vec{r}, t)}.$$ 
This common velocity receives a shift from the force $\vec{F}_{s}$ acting on the $s$ species \cite{SC_93,SD_95}.  This  force may be an external one or it could also be due to intermolecular (pseudo)-potential interactions.  The pseudo-potential force within each species consists of an attractive (a) component , acting only on the first  Brillouin region (belt, for simplicity), and a  repulsive (r) one acting on both belts, whereas  the force between species (X) is short-ranged and repulsive:  
$$\vec{F}_s(\vec{r}, t) = \vec{F}^a_s(\vec{r}, t) + \vec{F}^r_s(\vec{r}, t)+\vec{F}^X_s(\vec{r},t)
$$ 
where 


\begin{eqnarray}\label{FORCE}
\vec{F}^a_s(\vec{r}, t) &=& -G^a_s \Psi_s(\vec{r},t) \sum_{i \in belt 1} w_i \Psi_s(\vec{r}_{i},t) \vec{c}_{i}   \nonumber\\ 
\vec{F}^r_s(\vec{r}, t) &=&  - G^r_{s} \Psi_s(\vec{r},t) \sum_{i \in belt 1} p_{i} \Psi_s(\vec{r}_{i},t) \vec{c}_{i}  - G^r_{s} \Psi_s(\vec{r},t) \sum_{i \in belt 2} p_{i} \Psi_s(\vec{r}_{i},t) \vec{c}_{i}  \\ 
\vec{F}_s^X (\vec{r},t) &=& - \frac{1}{(\rho_0^{(s)})^2}\rho_s (\vec{r},t) \sum_{s' \neq s}\sum_{i \in belt 1}   G_{s s'}w_i \rho_{s'}(\vec{r}_i,t) \vec{c}_i \nonumber.
\end{eqnarray}

In the above, the groups 'belt $1$' and 'belt $2$'  refer to the first 
and second Brillouin zones in the lattice and $\vec{c}_{i}$, $p_{i}, w_i$ are  the  corresponding discrete speeds and associated weights (see figure \ref{fig:2} and table \ref{T1}).   Apart from a normalization factor, these correspond to the values given in \cite{Shan07,Sbragaglia07}. Also, $G_{ss'}=G_{s's}$, $s' \ne s$, is the cross-coupling between species, $\rho_0$ a reference  density to be defined shortly and, finally, $\vec{r}_{i} = \vec{r}+\vec{c}_{i}$  are the displacements along the $\vec{c}_{i}$ velocity vector. These interactions are sketched in Figure \ref{fig:1} for the case of a two component fluid (say species A and B).  Note that positive (negative) $G$ code for repulsion (attraction) respectively.  This  model is reminiscent of the potentials used to investigate arrested phase-separation  and structural arrest in charged-colloidal systems, and also bears similarities to  the NNN (next-to-nearest-neighbor) frustrated lattice spin models \cite{seth91,seth92,colloids05,Sciortino04}.  As compared with lattice spin models, in our case a high lattice connectivity is  required  to ensure compliance with macroscopic non-ideal hydrodynamics, particularly the isotropy of potential  energy interactions, which lies at the heart of the complex rheology to be discussed in this work.  To this purpose, the first belt is discretized with $9$ speeds, while the second   with $16$, for a total of $b=25$ connections (including rest-particles, for normalization purposes).  The weights are chosen in such a way as to fulfill the following normalization constraints: 
\be\label{NORM0}
w_0+\sum_{i \in belt 1} w_i = p_0+ \sum_{i \in belt 1} p_{i} + \sum_{i \in belt 2} p_{i} = 1
\ee
\be
\sum_{i \in belt 1} w_i c_{ix}^2 = \sum_{i \in belt 1} p_{i} c_{ix}^2 + \sum_{i \in belt 2} p_{i} c_{ix}^2= c_S^2.
\ee 
with $c_S^2=1/3$ the lattice sound speed and $w_0$ and $p_0$ the weights associated to the velocity at rest. All the weights take the values illustrated in Table \ref{T1}. The set of discrete speeds and corresponding weights are such as to recover 4th order isotropy for the interactions running on the first belt and 8th order isotropy for those extending over the second one. This choice is naturally patterned after reference \cite{Shan07,Sbragaglia07}, although  different options might be available.  The pseudo-potential $\Psi_s(\rho_s)$ is taken in the form originally suggested by Shan \& Chen \cite{SC_93,SC_94}
\begin{equation}
\label{PSI}
\Psi(\rho_s) = \rho_{0}^{(s)} (1-e^{-\rho_s/\rho_{0}^{(s)}}), 
\end{equation}
where $\rho_{0}^{(s)}$ marks the density value above which non ideal-effects come into play for species $s$. 
For the sake of simplicity, in the sequel we shall take a common value 
for all species, $\rho_{0}^{(s)} = \rho_0$.

\begin{table}[!h]
\begin{center}
\begin{tabular}{l l}
\hline
 \qquad \qquad \qquad Forcing Weigths (for $\vec{F}^r_s$) \\
\hline
  $p_{i}  = 247/420\;     $&$ \quad i = 0$\\
  $p_{i}  = 4/63\;     $&$ \quad i = 1,4$\\
  $p_{i}  = 4/135 \;   $&$ \quad i = 5,8$\\
  $p_{i}  = 1/180 \;   $&$ \quad i = 9,12$\\
  $p_{i}  = 2/945 \;   $&$ \quad i = 13,20$\\
  $p_{i}  = 1/15120 \; $&$ \quad i = 21,24$\\
\hline 
\end{tabular} 

\vspace{.2in}

\begin{tabular}{l l} 
\hline
 \qquad \qquad \qquad Forcing Weights (for $\vec{F}^a_s$ and $\vec{F}^X_s$) \\
\hline
  $w_{i}  = 4/9  $&$\quad i = 0$\\
  $w_{i}  = 1/9  $&$\quad i = 1,4$\\
  $w_{i}  = 1/36 $&$\quad i = 5,8$\\
\hline 
\end{tabular} 

\end{center}
\caption{\label{T1}\small{Links and weights of the two belts, $25$-speeds lattice \cite{SHAN06,Sbragaglia07} for all  interactions sketched in equations  (\ref{FORCE}). The first belt lattice velocities are indicated with $i=1...8$ while the second belt ones with $i=9...24$ (see also figure \ref{fig:2} for a sketch). $p_{i}$ or $w_i$ is indicating the weight associated with the $i$-th velocity in the various interactions. The weights associated to the velocity at rest, $w_0$ and $p_0$, are chosen to enforce  a unitary normalization (\ref{NORM0}).}}
\end{table}

\begin{figure}\label{fig:2}
\includegraphics[scale=0.4]{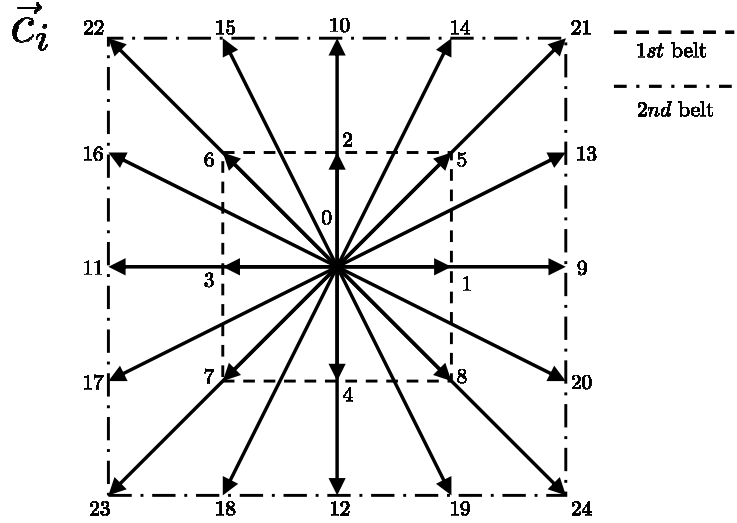}
\caption{The discrete $25$-speed lattice. Both belts are illustrated with the corresponding velocities.}
\label{fig:2}
\end{figure}

\section{Macroscopic equations}

The set of macroscopic equations associated with our kinetic model consists of the continuity  equations, one for each component separately, plus an equation of motion for total fluid momentum.  
Under the assumption of the same characteristic time scale for all the components $\tau_s=\tau$,  
\footnote[1]{Whenever timescales are different, the  characteristic time maps directly into an effective relaxation time. For the two species (A,B) system this takes the form $\bar \tau= \frac{\rho_A \tau_A+\rho_B \tau_B}{\rho}$. Similar  readjustments need to be used in the total barometric velocity}  these equations read as follows: 
\begin{equation}\label{eq:diff}
\partial_t \rho_s + \partial_a (\rho_s u_a) = \partial_a J_{sa}   
\end{equation}
\begin{equation}\label{EQMOM}
\partial_t (\rho u_a) + \partial_b (\rho u_a u_b ) = -\partial_b (c_S^2 \rho +\sigma_{ab})  
+ \sum _{s} F_{sa} = -\partial_{b} (P_{ab}-\sigma^{(visco)}_{ab})
\end{equation}
where $\rho=\sum_{s} \rho_s$ is the total density, $\vec{u} = \sum_s \rho_s \vec{u}_s/\rho$ is the baricentric (total) fluid  velocity, ${F}_{sa}$ the $a$-th component of the force acting on specie $s$  and $\sigma^{(visco)}_{ab}$ the dissipative component of the momentum-flux tensor.  The diffusive current in (\ref{eq:diff}) is given by 
\begin{equation}
\label{DIFFCUR}
J_{sa}=c_S^2 \left( \tau-\frac{1}{2 } \right) \left( \partial_a \rho_s -\frac{\rho_s}{\rho}\partial_a \rho \right) 
-\tau   \left( F_{sa} -\frac{\rho_{s}}{\rho} \sum_{s^{\prime}}F_{s^{\prime} a} \right). \end{equation}
Central to this analysis is the momentum-flux tensor, defined as the sum of a kinetic  component plus an interaction term:
\begin{equation}
P_{ab}=P^{kin}_{ab}+P^{int}_{ab}
\end{equation}
where
\begin{equation}
\label{PKIN}
P^{kin}_{ab} = \sum_{is} f_{is} c_{ia} c_{ib} 
\end{equation}
plus the interaction component, $P^{int}_{ab}$, defined by the condition:
\be\label{INTP}
\partial_b P^{int}_{ab} = -\sum_s F_{sa}.
\ee
Taylor expansion of the forcing terms will allow for a direct computation of $P^{int}_{ab}$ and the diffusion currents \cite{SD_95,SD_96}. It has to be noted that relation (\ref{INTP}) can also be directly satisfied on the lattice using the idea developed in a recent paper by Shan \cite{Shan08}, thus leading to more refined computational results for the momentum equation.

\subsection{Two component fluid}

The picture simplifies significantly for the case of a two-component fluid (say $A$ and $B$).  When the distribution functions $f_{iA},f_{iB}$ are close to the equilibrium, the kinetic part of the pressure tensor takes the following form:
\begin{equation}
\label{PBULKID}
P^{kin}_{ab} = (\rho_A+\rho_B) c_S^2 \delta_{ab}+K_{ab}^{(\tau)}
\ee
\be\label{PKINTAU}
K_{ab}^{(\tau)} =c_S^4 \frac{\rho_A \rho_B}{\rho} \left(\tau-\frac{1}{2} \right)^2\left(\frac{\partial_a \rho_A}{\rho_A}-\frac{\partial_a \rho_B}{\rho_B} \right)\left(\frac{\partial_b \rho_A}{\rho_A}-\frac{\partial_b \rho_B}{\rho_B} \right).
\end{equation}
where we recognize an ideal part $(\rho_A+\rho_B) c_S^2 \delta_{ab}$ plus some extra $\tau$ dependent terms. The origin of these terms will be elucidated in the section devoted to transport properties when we will detail the calculations of the surface tension coefficients across curved interfaces. Upon Taylor expanding \cite{Sbragaglia07} up to the fourth order the forcing terms  in the momentum equation, the interaction terms of the pressure tensor can be recast into the following form
\begin{equation}
\label{PBULKINT}
P^{int}_{ab} =  \left( c_S^2\frac{G_{A1}}{2} \Psi_A^2 + c_S^2\frac{G_{B1}}{2} \Psi_B^2 +c_S^2 g_{AB} \rho_A \rho_B  + c_S^4 \Pi \right) \delta_{ab}-\Gamma_{ab}
\end{equation}
\begin{equation}
\label{PSURF}
\Pi =  \sum_{s=A,B}G_{s2}\left( \frac{1}{4} |{\bm \nabla} \Psi_s|^2 + \frac{1}{2} \Psi_s \Delta \Psi_s \right)+  \frac{g_{AB}}{2} \left( \rho_A \Delta \rho_B + \rho_B \Delta \rho_A + {\bm \nabla} \rho_A  \cdot {\bm \nabla} \rho_B \right) 
\end{equation}
\begin{equation}
\label{P2}
\Gamma_{ab} = \frac{1}{2}c_S^4 \left( G_{2A} \partial_{a} \Psi_A \partial_{b} \Psi_A + G_{2B} \partial_{a} \Psi_B \partial_{b} \Psi_B +  g_{AB} (\partial_{a} \rho_A  \partial_{b} \rho_B + \partial_{a} \rho_B  \partial_{b} \rho_A) \right).
\end{equation}
In the above, we have set 
$$g_{AB} \equiv G_{AB}/\rho_0^2$$ 
and introduced the effective couplings 
\be\label{EFFECTIVE}
G_{s1}=G_s^a+G_s^r \hspace{.2in} G_{s2}=G_s^a+\frac{12}{7} G_s^r \hspace{.2in}  s=A,B.
\ee
From these expressions, we note that the presence of the second-neighbor repulsive layer allows a separate control of the equilibrium (equation of state, i.e. terms proportional to $c_S^2$) and  transport properties (surface tension, i.e. terms proportional to $c_S^4$).   For the diffusive current, we can Taylor expand the forcing terms up to the second order to obtain
\begin{equation}
J_{sa} = \sum_{s'} D_{ss'}(\rho_A,\rho_B) \partial_a \rho_{s'} \hspace{.3in} s,s'=A,B
\label{eq:diff}
\end{equation}
where the (non-linear) diffusion coefficients are given by:
\begin{equation}
D_{AA}=c_S^2 \left(\frac{\rho_B}{\rho}\left( \tau -\frac{1}{2} \right) + \frac{ \tau}{\rho}  (G_{A1} \rho_B \Psi_A\Psi'_A-g_{AB}\rho_B \rho_A) \right)
\end{equation}
\begin{equation}
D_{BB}=c_S^2\left( \frac{\rho_A}{\rho}\left( \tau -\frac{1}{2} \right) + \frac{ \tau}{\rho}(  G_{B1} \rho_A \Psi_B\Psi'_B-g_{AB}\rho_A \rho_B) \right)
\end{equation}
These are nothing but equations (26)-(29), already discussed in a earlier paper by Shan \& Doolen \cite{SD_95}. 
The above expressions indicate that the intra-species mass flow consists of an  internal component, proportional to the density of the other species, and a force-induced component, proportional to the intermolecular couplings \cite{SD_95,SD_96}.  Note that the latter does not vanish even in the limit of zero inter-species  interactions, $g_{AB} \rightarrow 0$.   The following reciprocity relations:
\begin{equation}
\label{RECIP}
D_{AB}=-D_{BB},\;\;\;D_{BA}=-D_{AA}.
\end{equation}
secure conservation of the total density. The continuum-time limit $\tau \gg \frac{1}{2}$ is thus characterized by
\begin{equation}
D_{AA} \rightarrow c_S^2 \tau \left(\frac{\rho_B}{\rho} + \frac{1}{\rho}  (G_{A1} \rho_B \Psi_A\Psi'_A-g_{AB}\rho_B \rho_A) \right)
\end{equation}
\begin{equation}
D_{BB} \rightarrow c_S^2 \tau \left( \frac{\rho_A}{\rho}+\frac{1}{\rho}(  G_{B1} \rho_A \Psi_B\Psi'_B-g_{AB}\rho_A \rho_B) \right)
\end{equation}
with the relaxation properties factorizing outside. 
It is therefore natural and convenient to introduce a $\tau$-dependent parameter 
\be\label{THETA}
\theta(\tau)=\frac{\tau-\frac{1}{2}}{\tau}
\ee
measuring the importance of discrete-time effects in the macroscopic equations. 
Clearly, in the continuum time limit $\theta(\tau) \rightarrow 1$, while  for
$\tau \rightarrow 1/2$ we have $\theta(\tau)=0$ (in terms of lattice Boltzmann fluids this is a dissipation free limit \cite{Gladrow00}). The diffusion coefficients can thus be recast into the following form
\begin{equation}
D_{AA}= \tau c_S^2 \left(\theta(\tau) \frac{\rho_B}{\rho} + \frac{1}{\rho}  (G_{A1} \rho_B \Psi_A\Psi'_A-g_{AB}\rho_B \rho_A) \right)
\end{equation}
\begin{equation}
D_{BB}= \tau c_S^2 \left(\theta(\tau) \frac{\rho_A}{\rho}+\frac{1}{\rho}(  G_{B1} \rho_A \Psi_B\Psi'_B-g_{AB}\rho_A \rho_B) \right).
\end{equation}

\section{Equilibrium Properties}

In this section we study the main equilibrium properties of the model previously introduced  whenever stable interfaces between the two fluids set in. 
To this purpose, we will focus on a one-dimensional problem, where inhomogeneities  in the density profiles develop only across a single coordinate, say $x$.  It proves expedient to start with the case of two components with mutual density repulsion (i.e. equation (\ref{FORCE}) with $\vec{F}_s^{r}=0$, $\vec{F}_s^{a}=0$ ), where  an {\it exact} matching with a free-energy functional can be achieved in the continuum limit, i.e. when  the discrete lattice effects are negligible. This allows us to envisage efficient strategies to describe the bulk equilibrium properties in special situations where all pseudo-potentials interactions are included 
(i.e. equation (\ref{FORCE}) with all the interactions on).

\subsection{Multicomponent Model with pure Density Repulsion}

At equilibrium, the relevant properties of the interfaces emerging from the
separation of the fluids can be obtained by imposing a constant diffusion 
current and a constant pressure all across the interface (zero net flow can safely be assumed). This yields:

\begin{equation}\label{EQCONT2}
\tau c_S^2 \left(\frac{\rho_B}{\rho}\theta(\tau) - \frac{1}{\rho}  (g_{AB}\rho_B \rho_A) \right)\partial_x \rho_A- \tau c_S^2\left( \frac{\rho_A}{\rho} \theta(\tau) - \frac{1}{\rho}(g_{AB}\rho_A \rho_B) \right)\partial_x \rho_B=J_0
\end{equation}
\begin{equation}\label{EQMOM2}
P_{xx} = c_S^2 \rho_A + c_S^2 \rho_B + c_S^2 g_{AB} \rho_A \rho_B  + c_S^4 \frac{g_{AB}}{2 } \left( \rho_A \partial_{xx} \rho_B + \rho_B \partial_{xx} \rho_A -  \partial_{x} \rho_A \partial_x \rho_B \right)+K_{xx}^{(\tau)}=P_0
\end{equation}
where $P_0$ is the constant value of the pressure across the interface 
and $J_0$ is the constant diffusion current predicted by the single component continuity equation. For simplicity we have not expanded the extra $\tau$ dependent terms ($K_{xx}^{(\tau)}$) of the kinetic pressure tensor (\ref{PKINTAU}).  Since $J_0=0$ in the bulk phases ($\partial_x \rho_{A,B}=0$), one concludes that    $J_0=0$ everywhere.  Next, we observe that the equation (\ref{EQCONT2}) can be recast  in the form of a differential equation relating the values of the two densities at each spatial location:
$$
\frac{d \rho_A}{d \rho_B}=\frac{\rho_A \rho^{(\tau)}_g-\rho_A \rho_B}{\rho_B \rho^{(\tau)}_g-\rho_A \rho_B}.
$$
In the above, we have defined 
\begin{equation}\label{RHOGTAU}
\rho^{(\tau)}_g=\frac{\theta(\tau)}{g_{AB}} 
\end{equation}
as a characteristic density  depending both on the relaxation properties in $\theta(\tau)$  and on the intermolecular coupling $g_{AB}$, above which inter-species repulsion becomes dominant. At the spatial location where $\rho_A=\rho_B$, we also have $\partial_{x} \rho_A=-\partial_{x} \rho_B$ because  of the symmetry of the system upon the interchange $\rho_A \leftrightarrow \rho_B$.  Equation (\ref{EQCONT2}) also shows that, at this location, $\rho_A=\rho_B=\rho^{(\tau)}_g$.  By integrating the previous differential equation backward and forward in density space, starting from the point where $\frac{d \rho_A}{d \rho_B}=-1$, it is possible to construct the manifold of density pairs  $(\rho_A,\rho_B)$ obeying the condition of zero mass flow.  For the specific case in point, these equations can be solved exactly, leading to the following relation
$$
\frac{\rho_A}{\rho_B}=exp \left( {{(\rho_A-\rho_B)/\rho^{(\tau)}_g}} \right).
$$
Obviously, this relation is fulfilled by the trivial solution $\rho_A=\rho_B$; owing to the non-linearity of the above equations, non trivial solutions are expected beyond a critical value of $g_{AB}$. These identify with the bulk densities once separation between the fluids has occurred.

Since we have neglected higher order terms in the Taylor-expansion yielding the diffusive current,  this relationship is not expected to hold uniformly across the interface. However, it can be be regarded as an excellent approximation 
to compute the bulk densities after separation of the two fluids.  To this end, we note that, out of the full set of pair densities, $(\rho_A,\rho_B)$ belonging to the density manifold, only one is compatible with the condition of equilibrium. Mechanical equilibrium, as obtained by imposing a constant pressure  tensor across the interface, equation (\ref{EQMOM2}), cannot serve as a selection criteria, because of the invariance under the interchange $\rho_A \leftrightarrow \rho_B$.  The two values of the bulk densities can however be fixed by imposing the total average density  in the numerical simulations $\langle \rho_A+\rho_B \rangle=\langle \rho \rangle$.  This provides a system of two equations determining the two bulk densities:
\begin{equation}\label{FIXING}
\begin{cases}
\frac{\rho_A}{\rho_B}=exp \left( {{(\rho_A-\rho_B)/\rho^{(\tau)}_g}} \right)\\
\rho_A+\rho_B=\langle \rho \rangle.
\end{cases}
\end{equation} 
Once the bulk densities have been fixed, the momentum equation (\ref{EQMOM2}), consistently with the  higher order in the Taylor expansion for the density equation (\ref{EQCONT2}), would allow to reconstruct the profiles across the interface.  Such technical construction will make the object of a forthcoming paper. \\

\subsection{Free-energy procedure}

\begin{figure*}[t!]
\begin{minipage}[t!]{16cm}
\includegraphics[width=16cm,keepaspectratio]{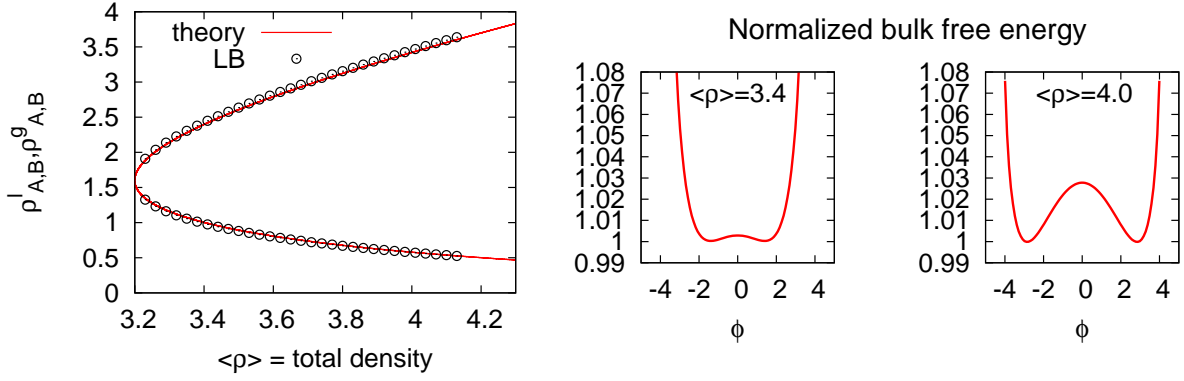}
\end{minipage}
\hfill
\begin{minipage}[t!]{16cm}
\caption{\label{fig:3} Equilibrium bulk densities in the multicomponent Shan-Chen model with simple density-density repulsion (equation (\ref{FORCE}) with $\vec{F}_s^{r}=0$, $\vec{F}_s^{a}=0$). In this case we have chosen $\rho_g^{(\tau)}=1.6$ in (\ref{RHOGTAU})  by setting $g_{AB}=0.345$, $\rho_0=1$ and $\tau=1.116071$. We have then simulated a $1d$ interface between two components at varying the total averaged density $\langle \rho \rangle$ in different numerical simulations. The numerical results are successfully compared with the prediction coming from the minimization of the free energy (\ref{FREE2}) and well approximated by equations (\ref{TAYLORRES}). In the right figures we also show the typical profiles of the bulk free energies arising in the theory behind the simulations.  For simplicity the bulk free energy has been normalized to minus a unit value at the two minima. All results are given  in lattice Boltzmann units (LBU).}
\end{minipage}
\end{figure*}

In order to better elucidate the mechanism fixing the bulk densities in the phase separation process, we can also resort to a direct {\it exact} link with a free energy functional in the continuum limit,  where all discrete lattice effects disappear.  We begin by considering a free-energy density in the form
\begin{equation}\label{FFF}
{\cal L}(\rho_A,\rho_B)= c^2_S \rho_A \log  \rho_A +c^2_S \rho_B \log  \rho_B +c_S^2 g_{AB} \rho_A \rho_B -c_S^4 \frac{g_{AB}}{2} {\bm \nabla} \rho_A \cdot  {\bm \nabla} \rho_B . 
\end{equation}
This consists of the sum of two ideal free-energy densities  ($ c_S^2 \rho_{A,B} \log \rho_{A,B}$) plus an interaction term. It has to be stressed that the terms proportional to $c_S^2 g_{AB}$ in front of the interacting terms should by no means be related to the fluid temperature, as they simply disappear upon a suitable choice of the lattice forcing weights \cite{SHAN06}.  On the other hand, the term proportional to $c_S^2$ in front of the ideal parts ($\sim \rho_{A,B} \log \rho_{A,B}$)  plays the role of a global reference temperature.  This is of no relevance for the present athermal case, but may become important for  generalizations involving temperature fluctuations \cite{NOI09}, where internal energies need to be introduced.  As to the free-energy in (\ref{FFF}), it is readily checked that the bulk contribution
$$
f_b(\rho_A,\rho_B)= c^2_S \rho_A \log  \rho_A +c^2_S \rho_B \log  \rho_B +c_S^2 g_{AB} \rho_A \rho_B 
$$
correctly reproduces the bulk pressure:
$$
P_b(\rho_A,\rho_B)=\rho_A \frac{\partial f_b}{\partial \rho_A}+\rho_B \frac{\partial f_b}{\partial \rho_B}-f_b=c_S^2 (\rho_A+\rho_B)+c_S^2 g_{AB} \rho_A \rho_B
$$
that is the generalization of the standard Legendre's relation $P_b(\rho) = \rho \frac{\partial f(\rho)}{\partial \rho}-f(\rho)$   connecting the free-energy to the bulk pressure of a single-component fluid.  In order to preserve both densities separately, we next introduce two  Lagrange multipliers, say $\lambda_A$ and $\lambda_B$, thus leading to the following 
constrained free-energy density: 
\begin{equation}\label{FREE}
{\cal L}(\rho_A,\rho_B)=f_b(\rho_A,\rho_B)-c_S^4 \frac{g_{AB}}{2} {\bm \nabla} \rho_A \cdot {\bm \nabla} \rho_B -\lambda_A \rho_A-\lambda_B \rho_B.
\end{equation}
Variations of this constrained free-energy with respect to $\rho_A$ and $\rho_B$ delivers the following two Euler-Lagrange equations:
\be
\begin{cases}
\frac{\partial {\cal L}}{\partial \rho_{A}}  -\partial_{\alpha} \left(  \frac{\partial {\cal L}}{\partial (\partial_{\alpha} \rho_{A})} \right)=0 \\
\frac{\partial {\cal L}}{\partial \rho_{B}}  -\partial_{\alpha} \left( \frac{\partial {\cal L}}{\partial (\partial_{\alpha} \rho_{B})} \right) =0. 
\end{cases}
\ee
Based on (\ref{FFF}), these yield:
\begin{equation}\label{SETSETSET}
\begin{cases}
\frac{d f_b}{d \rho_A}+c_S^4 \frac{g_{AB}}{2}  \partial_{xx} \rho_B=\lambda_A\\
\frac{d f_b}{d \rho_B}+c_S^4 \frac{g_{AB}}{2}  \partial_{xx} \rho_A=\lambda_B.
\end{cases}
\end{equation}
Upon multiplying the first equation by $\frac{\partial \rho_A}{\partial x}$ and the second  by $\frac{\partial \rho_B}{\partial x}$ we can then integrate  between the bulk region ($x=0$) and  a generic  interface location ($x$). In this way,  we obtain
\begin{equation}\label{2NOETHER}
\begin{cases}
- \left. c_S^2 \rho_A \right|_0^{x}- \left. c_S^2 g_{AB}\rho_A \rho_B \right|_{0}^{x} + c_S^2 g_{AB} \int_{0}^{x}\rho_B \partial_{y} \rho_A dy+c_S^4 \frac{g_{AB}}{2} \int_{0}^{x}(\partial_y \rho_A)(\partial_{yy} \rho_B) dy-c_S^4 \frac{g_{AB}}{2} \rho_A\partial_{xx} \rho_B=0  \\
- \left. c_S^2 \rho_B \right|_0^{x}- \left. c_S^2 g_{AB}\rho_A \rho_B \right|_{0}^{x} + c_S^2 g_{AB} \int_{0}^{x}\rho_A \partial_{y} \rho_B dy+c_S^4 \frac{g_{AB}}{2} \int_{0}^{x}(\partial_y \rho_B)(\partial_{yy} \rho_A) dy-c_S^4 \frac{g_{AB}}{2} \rho_B\partial_{xx} \rho_A=0
\end{cases}
\end{equation}
where $...\left. \right|_{0}^{x}$ represents the variation between $0$ (bulk) and $x$ (interface location) of the desired observable. The above equations represent the conserved currents associated with the two Lagrange multipliers and they can be linked directly into the constant pressure tensor and diffusion current at equilibrium. In fact, by summing both equations in (\ref{2NOETHER}) we obtain 
\be
c_S^2 (\rho_A+\rho_B)+c_S^2 g_{AB} \rho_A \rho_B+\frac{c_S^4 g_{AB}}{2} \left( \rho_B\partial_{xx} \rho_A + \rho_A\partial_{xx} \rho_B-\partial_x \rho_A\partial_{x} \rho_B \right) =\mbox{const.}
\ee
that is reminiscent of (\ref{EQMOM2}) upon neglecting $K_{xx}^{(\tau)}$. 
Similarly, upon applying the $x$ derivative to both equations in (\ref{2NOETHER}) 
and then 
multiplying the first equation by $\rho_B$ and the second by $\rho_A$ we can finally subtract the two contributions to get 
$$
(\rho_B\partial_x \rho_A-\rho_A\partial_x \rho_B) - g_{AB}\rho_B \rho_A (\partial_x \rho_A-\partial_x \rho_B)+{\cal O}(\partial^3) =0
$$
that is delivering  the condition of a zero diffusion current, as  given in (\ref{EQCONT2}) with $J_0=0$, in the limit $\tau \gg \frac{1}{2}$, i.e. $\theta(\tau) \rightarrow 1$.  Thus, in the free-energy formalism, both conservations descend from the same single scalar. The free-energy formalism permits to recast the continuity equations in terms of  the gradients of the chemical potentials 
$
\mu_s=\frac{\partial {\cal L}}{\partial \rho_s}.
$
More specifically:
\begin{equation}
\partial_t \rho_{A} + \partial_a (\rho_{A} u_a) = \partial_{a} (M(\rho_A,\rho_B) \partial_a  (\mu_{A}-\mu_{B} ))
\end{equation}
\begin{equation}
\partial_t \rho_{B} + \partial_a (\rho_{B} u_a) = \partial_{a} (M(\rho_A,\rho_B) \partial_a  (\mu_{B}-\mu_{A} ))
\end{equation}
where the mobility $M(\rho_A,\rho_B)$ is given by $M(\rho_A,\rho_B)=\tau \frac{\rho_A \rho_B}{\rho}$. The above form of the continuity equation explicitly shows that mass diffusion is triggered  by an unbalance of the local chemical potentials, so that equilibrium is attained whenever $\mu_A=\mu_B$.  So much for the continuum picture. \\
For a finite value of $\tau$, an exact matching between momentum and continuity equations starting from  continuum free-energy functional (\ref{FREE}) is not so straightforward and more elaborate arguments are necessary.  It is however possible  to fix the bulk densities by introducing the following $\tau$-dependent functional
\begin{equation}\label{FREE2}
{\cal L}^{(\tau)}(\rho_A,\rho_B)=f^{(\tau)}_b(\rho_A,\rho_B)-c_S^4\frac{g^{(\tau)}_{AB}}{2} {\bm \nabla} \rho_A \cdot {\bm \nabla} \rho_B -\lambda_A\rho_A-\lambda_B \rho_B
\end{equation}
\begin{equation}\label{FREE2}
f^{(\tau)}_b(\rho_A,\rho_B)=c^2_S \rho_A \log  \rho_A +c^2_S \rho_B \log  \rho_B +c_S^2 g^{(\tau)}_{AB} \rho_A \rho_B 
\end{equation}
where 
$g^{(\tau)}_{AB}=\frac{g_{AB}}{\theta(\tau)}$ is the effective coupling  renormalized by lattice discreteness effects (note that this is exactly the inverse of the reference density $\rho^{(\tau)}_g$ introduced earlier on).  We note that in the long-time limit $\tau \gg 1/2$ ($\theta(\tau) \rightarrow 1$), we have
$$
g^{(\tau)}_{AB} \rightarrow g_{AB}
$$
thus reproducing the continuum value. This $\tau$ dependence of the effective coupling reflects into an analogue dependence of the bulk densities.  The bulk minimization with respect of  $\rho_A$ and $\rho_B$, along the same lines as for the continuum case, leads to the following bulk equations (the same procedure leading to (\ref{SETSETSET}), with $\partial_{xx} \rho_{A,B}=0$)
\begin{equation}\label{manifold}
\begin{cases}
\frac{d f_b^{(\tau)}}{d \rho_A}+c_S^2 g^{(\tau)}_{AB} \rho_B=\lambda_A\\
\frac{d f_b^{(\tau)}}{d \rho_B}+c_S^2 g^{(\tau)}_{AB} \rho_A=\lambda_B.
\end{cases}
\end{equation}
The symmetry under the interchange $\rho_A \leftrightarrow \rho_B$  imposes $\lambda_A=\lambda_B$. 
By subtracting the second from the first equation in (\ref{manifold}) we obtain again the relation (\ref{FIXING}), 
thus showing that the manifold of $(\rho_A,\rho_B)$  minimizing the bulk free-energy is {\it the same as the one obtained by imposing a zero diffusion current}.  Here again, in order to single out a point of minimum, we have to specify the total mass in the system, as stated in (\ref{FIXING}). The procedure gains transparency by replacing the densities with their their local sum ($\rho=\rho_A+\rho_B$) and difference ($\phi=\rho_A-\rho_B$). The bulk-free energy functional takes then the following form:
\begin{equation}\label{FREE2}
f_b^{(\tau)}(\rho,\phi)=\frac{c_S^2}{2}(\rho+\phi)\log \left(\frac{\rho+\phi}{2} \right)+\frac{c_S^2}{2}(\rho-\phi)\log \left(\frac{\rho-\phi}{2} \right) -\frac{c_S^2}{4} g^{(\tau)}_{AB} (\rho^2-\phi^2)
\end{equation}
It can be checked that, at a given value of $\rho$, this expression presents a double well structure, as soon as $\rho \ge \rho^{(\tau)}_g$.  The two minima correspond to the two symmetric values of $\phi$ ($\pm \phi_0$) attained at equilibrium in the bulk phases.  To be noted that the presence of the two minima reminds of the 'double tangent description'  characterizing the minimization of a free-energy functional.  In this simple case, due to the symmetric structure of the problem, we are left with a symmetric  free energy and therefore the bulk densities can be directly extracted from those two minima.  By Taylor expanding the full set of equations (\ref{FIXING}) we obtain an analytical estimate of the solution for the two bulk densities ($\rho^l_{A,B},\rho^h_{A,B}$  where $l,h$ stands for low and high density), namely:
\be\label{TAYLORRES}
\begin{cases}
\rho^l_{A,B}=\frac{\langle \rho \rangle +\sqrt{30 (\rho^{(\tau)}_g)^2-6 \rho^{(\tau)}_g \sqrt{45 (\rho^{(\tau)}_g)^2-10 \rho^{(\tau)}_g \langle \rho \rangle}}}{2}\\
\rho^h_{A,B}=\frac{\langle \rho \rangle-\sqrt{30  (\rho^{(\tau)}_g)^2-6 \rho^{(\tau)}_g \sqrt{45 (\rho^{(\tau)}_g)^2-10 \rho^{(\tau)}_g \langle \rho \rangle}}}{2}
\end{cases}.
\ee
This approach has been validated against numerical simulations. The results, referring to the case $\rho_g^{(\tau)}=1.6$, $g_{AB}=0.345$, $\rho_0=1.0$ and  $\tau=1.116071$ in lattice Boltzmann units (LBU), are shown in figure \ref{fig:3}. We have simulated a $1d$ interface between two components at varying the total averaged density $\langle \rho \rangle$.  The numerical results compare satisfactorily with the theoretical predictions based on the minimization of  the free energy (\ref{FREE2}).  In the right panel of the same figure, also shown are typical profiles of the bulk free energies arising in the numerical study.


\subsection{Multicomponent Model with Self-Interactions}

\begin{figure*}[t!]
\begin{minipage}[t!]{16cm}
\includegraphics[width=10cm,keepaspectratio]{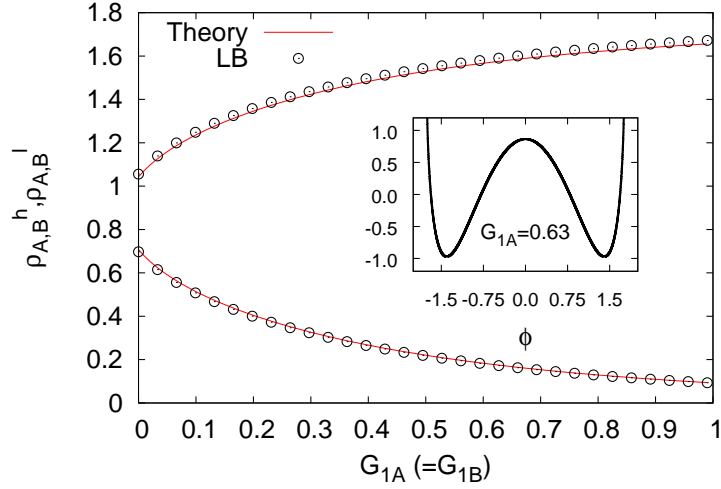}
\end{minipage}
\hfill
\begin{minipage}[t!]{16cm}
\caption{\label{fig:4} Bulk densities in the numerical simulations with repusion plus pseudo-potentials (i.e. equation (\ref{FORCE}) with $\vec{F}_s^{r}=0$). Both pseudo-potentials are chosen in the same way, $\Psi_{A,B}=\rho_0 (1-e^{-\rho_{A,B}/\rho_0})$ fixing $\rho_0=1.0$, $\langle \rho \rangle=1.75$, $\tau=1.0$, $g_{AB}=0.5785$. By changing the self-coupling parameters in such a way that $G_{A1}=G_{A2}$ we have computed the equilibrium bulk densities and also compared with the results coming from the minimization of the bulk (symmetric) free energy (\ref{BULKfinal}). In the inset figure we also show the typical profile of the bulk free energy arising in the theory behind the simulations at $G_{1A}=-0.63$.  For simplicity the bulk free energy has been normalized to minus a unit value at the two minima. All results are reported in LBU. }
\end{minipage}
\end{figure*}

Having covered the case with purely repulsive inter-species interactions, we next consider the more general situation in which intra-species (self) interactions are included (equation (\ref{FORCE}) with all interactions on).   In this general case, the condition of no mass diffusion ($J_0=0$) delivers:
\begin{equation}
\sum_{s=A,B} D_{As}(\rho_A,\rho_B) \partial_x\rho_{s}=\sum_{s=A,B} D_{Bs}(\rho_A,\rho_B) \partial_x \rho_{s}=0
\end{equation}
with
\begin{equation}
D_{AA}=c_S^2 \tau \left(\frac{\rho_B}{\rho}\theta(\tau) + \frac{1}{\rho}  (G_{A1} \rho_B \Psi_A\Psi'_A-g_{AB}\rho_B \rho_A) \right)
\end{equation}
\begin{equation}
D_{BB}=c_S^2 \tau \left( \frac{\rho_A}{\rho} \theta(\tau) + \frac{1}{\rho}(  G_{B1} \rho_A \Psi_B\Psi'_B-g_{AB}\rho_A \rho_B) \right)
\end{equation}
with the usual symmetries: $D_{AB}=-D_{BB},\;\;\;D_{BA}=-D_{AA}$.  Also, a constant ($P_0$) pressure tensor across the interface is required:
\begin{equation}\label{EQPSEUDO1}
P_{xx} = \left( c_S^2 \rho_A + c_S^2 \rho_B +  \frac{1}{2}c_S^2 G_{A1} \Psi_A^2 + 
\frac{1}{2}c_S^2 G_{B1} \Psi_B^2+c_S^2 g_{AB}\rho_A \rho_B + c_S^4 \Pi \right) - \Gamma_{xx}+K_{xx}^{(\tau)}=P_0
\end{equation}
\begin{equation}\label{EQPSEUDO2}
\Pi = \Sigma_{s=A,B}G_{s2}\left( \frac{1}{4} (\partial_x \Psi_s)^2 + \frac{1}{2} \Psi_s \partial_{xx} \Psi_s \right)+\frac{g_{AB}}{2} \left( \rho_A \partial_{xx} \rho_B + \rho_B \partial_{xx} \rho_A + \partial_x \rho_A \partial_x \rho_B \right) 
\end{equation}
\begin{equation}\label{EQPSEUDO3}
\Gamma_{xx} = \frac{c_S^2}{2} \left( G_{2A} \partial_{x} \Psi_A \partial_{x} \Psi_A + 
G_{2B} \partial_{x} \Psi_B \partial_{x} \Psi_B + 
g_{AB}  (\partial_{x} \rho_A  \partial_{x} \rho_B + \partial_{x} \rho_B  \partial_{x} \rho_A ) \right)
\end{equation}
with the various effective couplings already defined in (\ref{EFFECTIVE}) and $K_{xx}^{(\tau)}$  defined in (\ref{PKINTAU}). These two 'conserved' currents must be matched with the total mass in the system.  In the most general case, we expect two characteristic values of the sum of the two densities  in the two bulks, corresponding to the four unknowns $\rho^{l}_A,\rho^{h}_A$ and $\rho^{l}_B,\rho^{h}_B$.  One can resort again to a minimization procedure based on the following free-energy density
\begin{equation}\label{FREETOTAL}
{\cal L}(\rho_A,\rho_B)=f_b(\rho_A,\rho_B)+c_S^4 \frac{G_{A2}}{2} |{\bm \nabla} \Psi_A|^2 +c_S^4 \frac{G_{B2}}{2} |
{\bm \nabla} \Psi_B|^2 -  c_S^4 \frac{g_{AB}}{2} {\bm \nabla} \rho_A \cdot {\bm \nabla} \rho_B -\lambda_A \rho_A-\lambda_B \rho_B
\end{equation}
with the bulk contribution written as
$$
f_b(\rho_A,\rho_B)= c^2_S \rho_A \log  \rho_A +c^2_S \rho_B \log  \rho_B +c_S^2 g_{AB} \rho_A \rho_B+c_S^2\frac{G_{A1}}{2} \rho_A \displaystyle\int_{0}^{\rho_A} \frac{\Psi^2_A(\xi)}{\xi^2} d\xi +c_S^2\frac{G_{B1}}{2}\rho_B \displaystyle\int_{0}^{\rho_B} \frac{\Psi_B^2(\xi)}{\xi^2} d\xi. 
$$
Note that, like in the purely repulsive case, this matches the equilibrium properties  of our system in the limit $\tau  \gg 1/2 $, where lattice time discreteness can be ignored.  Moreover, due to the presence of the pseudo-potentials $\Psi$, in order to make the Shan-Chen  model compliant with such a kind of free energy, an extra-gradient term has to be added, as  described in a recent paper \cite{Sbragaglia09}.  Such extra-term is connected with variations of the pseudo-potentials across the interface   and, at least for the case of a single-component fluid, it can be shown to be negligible to practical purposes. \\ It is also worth noting that in the symmetric case  $G_{A1}=G_{B1}$ (the one analyzed later in the paper)  with the same pseudo-potential for both components $\Psi_A=\Psi_B$, we can use similar arguments  as described in the previous subsection.  In particular, we define the following $\tau$-dependent bulk free energy
\begin{equation}\label{BULKfinal}
f^{(\tau)}_b(\rho_A,\rho_B)= c^2_S \rho_A \log  \rho_A +c^2_S \rho_B \log  \rho_B +c_S^2 g_{AB}^{(\tau)} \rho_A \rho_B+c_S^2 \frac{G^{(\tau)}_{A1}}{2}\rho_A \displaystyle\int_{0}^{\rho_A} \frac{\Psi_A^2(\xi)}{\xi^2} d\xi +c_S^2 \frac{G^{(\tau)}_{B1}}{2} \rho_B \displaystyle\int_{0}^{\rho_B} \frac{\Psi_B^2(\xi)}{\xi^2} d\xi 
\end{equation}
with $G^{(\tau)}_{A1}=\frac{G_{A1}}{\theta(\tau)}$, $G^{(\tau)}_{A2}=\frac{G_{A2}}{\theta(\tau)}$ 
and look for its (symmetric) minima.  
In figure \ref{fig:4}, we show the comparison between the results
of minimization of this free energy and those by direct numerical simulations
with the usual pseudo-potential $\Psi=\rho_0 (1-e^{-\rho/\rho_0})$. 
The main parameters are $\rho_0=1.0$, $\langle \rho \rangle=1.75$, $g_{AB}=0.345$ and $\tau=0.69$ LBU, and different values of the self coupling parameters $G_{A1}=G_{A2}$.  Overall, satisfactory agreement is observed.

\section{Transport properties}

In this section we focus on the theoretical prediction of the surface tension of the two-component model. As previously discussed, this requires the correct identification of the off-diagonal  component of the momentum-flux tensor $P_{ab}$ .  For the sake of concreteness, we shall consider the simplest case of a one-dimensional stationary interface between the two fluids $A$ and $B$.  In view of equation (\ref{PBULKINT}), the surface tension is given by
\begin{equation}
\label{s1}
\sigma_{AB} = -\int_{flat} \Gamma_{xx} dx 
\end{equation}
where $\int_{flat}$ is a short hand notation for integration across a flat interface separating the two fluids and developing across $x$.  However, as pointed out by Shan \& Chen \cite{SC_94}, the time discretization induces an extra term on the r.h.s. of (\ref{s1})  and, given its importance for the actual computation of the surface tension, in the following we shall generalize their treatment to the case of a two-component fluid.  We start by writing the lattice kinetic equation for the total distribution function $g_i \equiv f_{iA}+f_{iB}$:
$$
g_{i}(\vec{r} + \vec{c}_i, t + 1)-g_{i}(\vec{r} , t )=-\frac{1}{\tau}[g_{i}(\vec{r} , t )-g_{i}^{(eq)}(\rho_A,\rho_B,\vec{u},\vec{F}_A,\vec{F}_B)]   
$$
where the total equilibrium is simply the sum of the two single-component equilibria
$$
g_{i}^{(eq)}=f_{iA}^{(eq)}(\rho_A ,  \vec{u}+\tau \vec{F}_{A}/\rho_A)+f_{iB}^{(eq)}(\rho_s ,  \vec{u}+\tau \vec{F}_{B}/\rho_B).
$$
By unrolling the full expressions of $f_{iA}^{(eq)}$ and $f_{iB}^{(eq)}$, we obtain:
\be\label{MAXTOT}
g_{i}^{(eq)}=w_i^{(eq)} \left(\rho+\rho \frac{{u}^{(eq)}_a {c}_{ia}}{c_S^2}+ \frac{({c}_{ia} {c}_{ib}-c_S^2 {\delta_{ab}})}{2 c_S^4}\left(\rho  {u}^{(eq)}_a {u}^{(eq)}_b  + \tau^2 \frac{{F}_{Aa}{F}_{Ab}}{\rho_A}+ \tau^2\frac{{F}_{Ba}{F}_{Bb}}{\rho_B}- \tau^2\frac{{F}_{a}{F}_{b}}{\rho} \right) \right)
\ee
where $F_a=F_{Aa}+F_{Ba}$ is the $a$-th component of the total force and
\be\label{DEFUEQ}
\vec{u}^{(eq)}=\vec{u}+\tau \vec{F}/\rho
\ee
is the total fluid velocity, including the shift due to the total force. To be noted that the term $ \tau^2 (\frac{{F}_{Aa}{F}_{Ab}}{\rho_A}+\frac{{F}_{Ba} {F}_{Bb}}{\rho_B}-\frac{{F}_a {F}_{b}}{\rho})$ 
is missing in the original paper by Shan \& Chen \cite{SC_94}, because these authors deal with 
a single-species fluid.  Next, following \cite{SC_94}, we estimate $\vec{u}^{(eq)}$ by general considerations holding at steady state.  For stationary solutions, we can assume no net mass transfer along any link  connecting two lattice sites, which implies $g_i(\vec{r}+ \vec{c_i}) = g_j(\vec{r})$, where $j$ is the mirror partner defined by the condition $\vec{c}_j= - \vec{c}_i$.  Under this constraint, one derives the relation \cite{SC_94}:
\begin{equation} \label{SC94}
-2 \rho \vec{u} = -\frac{1}{\tau}(\rho \vec{u} - \rho \vec{u}^{(eq)})
\end{equation}
which, combined with (\ref{DEFUEQ}), delivers:
$$
\rho \vec{u}^{(eq)}=\left(\tau-\frac{1}{2} \right) \vec{F}=\tau \theta(\tau) \vec{F}.
$$
This expression can then be used to evaluate the kinetic component of the pressure tensor (\ref{PKIN}),
$$
P^{kin}_{ab}=\sum_i g_{i} {c}_{ia} {c}_{ib}.
$$
By assuming $g_{i} \approx g_{i}^{(eq)}$, the term $\sum_i g_{i} {c}_{ia} {c}_{ib}$ delivers the following contribution:
\be\label{TAUCORR}
\sum_i g_{i}^{(eq)} {c}_{ia} {c}_{ib} =c_S^2 \rho \delta_{ab}+\tau^2 \theta^2(\tau) \frac{F_a F_b}{\rho}+\tau^2\left(\frac{F_{Aa} F_{Ab}}{\rho_A}+\frac{F_{Ba} F_{Bb}}{\rho_B}-\frac{F_a F_b}{\rho} \right)
\ee
where, as anticipated in the previous sections,  we recognize the ideal gas equation of state, plus extra $\tau$-dependent contributions stemming from the forcing terms.  This shows that discrete effects (both in time and space) introduce a correction to the surface tension, which  must be taken into account in order to compute the value of $\sigma_{AB}$.  We can now make use of the identity
\be\label{EQSUPPORT}
\tau^2\left(\frac{F_{Aa} F_{Ab}}{\rho_A}+\frac{F_{Ba} F_{Bb}}{\rho_B}-\frac{F_{a} F_{b}}{\rho} \right)=\tau^2 \frac{\rho_A \rho_B}{\rho} \left(\frac{F_{Aa}}{\rho_A}-\frac{F_{Ba}}{\rho_B} \right)\left(\frac{F_{Ab}}{\rho_A}-\frac{F_{Bb}}{\rho_B} \right).
\ee
Also, the condition of no mass-diffusion current (\ref{eq:diff}), gives: 
\be\label{NOMASSNO}
\left(\frac{F_{Aa}}{\rho_A}-\frac{F_{Ba}}{\rho_B} \right)=c_S^2 \theta(\tau) \left(\frac{\partial_a \rho_A}{\rho_A}-\frac{\partial_a \rho_B}{\rho_B} \right).
\ee
Inserting (\ref{EQSUPPORT}) together with (\ref{NOMASSNO})  into the rhs of (\ref{TAUCORR}), finally delivers
\be\label{FINALEXTRA}
\sum_i g_{i}^{(eq)} {c}_{ia} {c}_{ib} =c_S^2 \rho \delta_{ab}+\tau^2 \theta^2(\tau)\left[ \frac{F_{a} F_{b}}{\rho}+c_S^4 \frac{\rho_A \rho_B}{\rho} \left(\frac{\partial_a \rho_A}{\rho_A}-\frac{\partial_a \rho_B}{\rho_B} \right)\left(\frac{\partial_b \rho_A}{\rho_A}-\frac{\partial_b \rho_B}{\rho_B} \right) \right]
\ee
which is precisely the result reported in (\ref{PKINTAU}). 

In conclusion, the expression for the overall surface tension must
take into account the contribution of both the potential energy and the 
($\tau$-dependent) kinetic energy components of the pressure tensor:
\begin{equation}
\label{SURFACEFINAL}
\sigma_{AB} = -\int_{flat} \Gamma_{xx} dx+\int_{flat}K_{xx}^{(\tau)}   dx 
\end{equation}
with $\Gamma_{xx}$ and $K_{xx}^{(\tau)}$ stemming from the
interaction pressure tensor (equation (\ref{P2})) and the $\tau$-dependent part of 
the kinetic pressure tensor (equation (\ref{PKINTAU})), respectively. 

\begin{figure*}[t!]
\begin{minipage}[t!]{16cm}
\includegraphics[width=10cm,keepaspectratio]{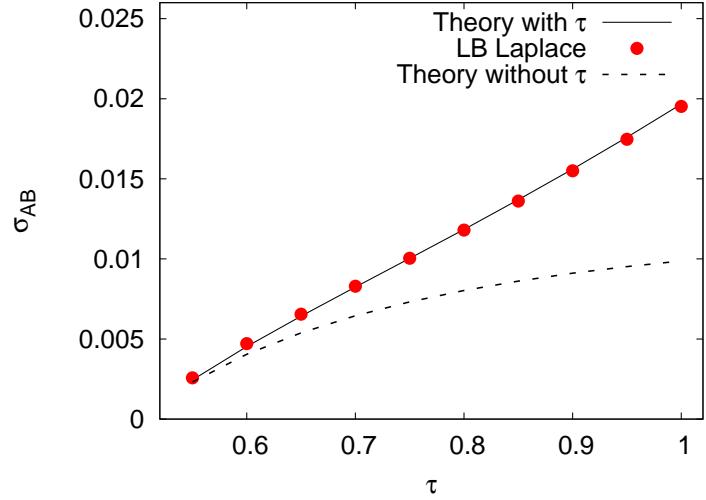}
\end{minipage}
\hfill
\begin{minipage}[t!]{16cm}
\caption{\label{fig:5}Surface tension in a model with pure particle-particle repulsion (equation (\ref{FORCE}) with $\vec{F}_s^{r}=0$, $\vec{F}_s^{a}=0$)  .  We have fixed $\rho^{(\tau)}_g=0.91$ in LBU as defined in equation (\ref{RHOGTAU})  and then varied $\tau$ in the lattice Boltzmann simulations (all the numerical values reported are in LBU). Results show the bare surface tension computed with the simple interaction pressure tensor and also with the $\tau$ corrections as described in (\ref{TAUCORR}). } 
\end{minipage}
\end{figure*}

The presence of the extra $\tau$ dependent terms has been checked against numerical simulations with pure repulsion (equation (\ref{FORCE}) with $\vec{F}_s^{r}=0$, $\vec{F}_s^{a}=0$ ), as shown in figure \ref{fig:5}.  We have fixed $\rho^{(\tau)}_g=0.91$ in LBU and varied $\tau$ in the  simulations.  The numerical results in figure \ref{fig:5} show the bare surface tension computed with and without  the $\tau$ corrections given in (\ref{TAUCORR}), as well as through the usual Laplace test, i.e..  by evaluating the difference between inner $P_{in}$ and outer $P_{out}$ equilibrium bulk pressure of two-dimensional droplets of radius $R$, and extracting the surface tension from the Laplace's relation:
$$
P_{in}-P_{out}=\frac{\sigma_{AB}}{R}.
$$
The results clearly indicate that the correction terms are essential 
to achieve quantitative agreement with the Laplace's values. 
To be noted that the $\tau$-dependence of the equilibrium component of the kinetic pressure tensor, rhs of equation (\ref{FINALEXTRA}), which stems from the shifted  velocity in the local equilibrium, disappears in the limit $\tau \rightarrow 1/2$. 

\subsection{Achieving vanishingly-low surface tension for finite relaxation times}

\begin{figure*}[t!]
\includegraphics[scale=0.75]{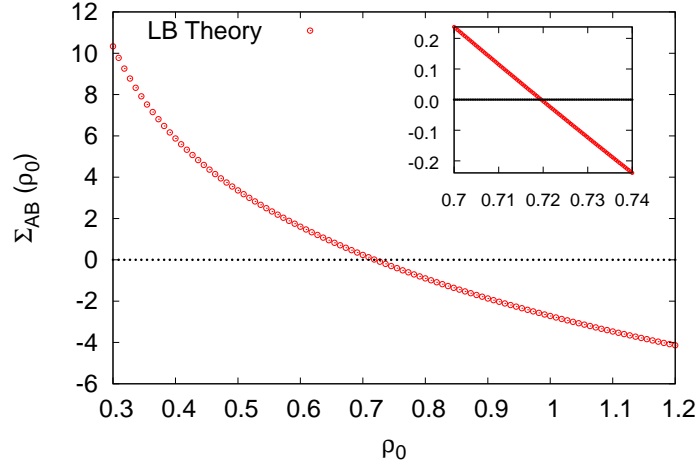}
\hfill
\caption{\label{fig:6} The function $\Sigma(\rho_0)$ as defined in (\ref{surften}) for the set of parameters $\langle \rho \rangle = 1.23$, $G_{AB}=0.405$, $G_{A2}=9.1714$, $G_{B2}=8.45714$.  The crossover lies at about $\rho_0 = 0.72$, as also evidenced in the inset.  The line at zero surface tension is reported as a visual guideline.  All results are reported in LBU.} 
\end{figure*}

\begin{figure*}[t!]
\includegraphics[scale=0.75]{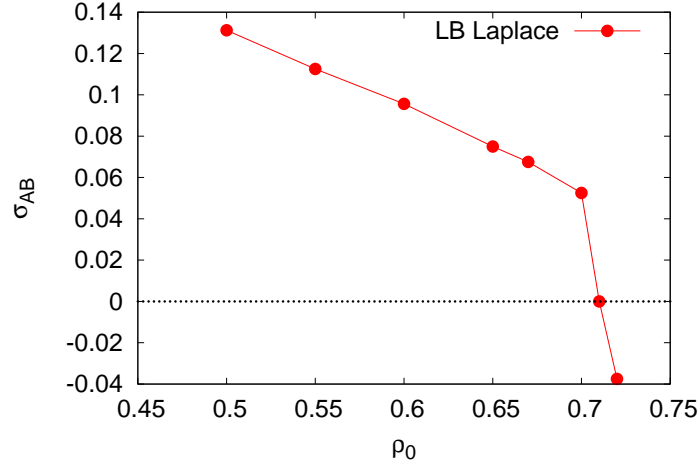}
\hfill
\caption{\label{fig:7} Surface tension as a function of $\rho_0$, as obtained by numerical simulations with the parameters $\langle \rho \rangle = 1.23$, $G_A^a = -15.0$, $G_A^r = 14.1$,    $G_B^a = -14.0$, $G_B^r = 13.1$, $G_{AB} = 0.405$. This produces in $G_{A2}=9.1714$, $G_{B2}=8.45714$ in (\ref{sigmaAB}). The crossover lies around $\rho_0 \sim 0.71$ LBU, in close agreement with the analytical estimate. The line at zero surface tension is reported for eye-guiding
purposes. All results are given in LBU.}
\end{figure*}

Going back to the general expression of the forcing terms (\ref{FORCE}) it is interesting to observe that, once the values of the $G$-couplings in the full model are fixed, we can still tune the surface tension by suitably changing $\rho_0$ in the model.  For a  fixed relaxation time $\tau$ (say $\tau=1$ LBU) this turns out to be a practical computational strategy to access the vanishing low surface-tension regime of interest for the simulation of micro-emulsions. Using the theory developed so far, we can now estimate the surface tension $\sigma_{AB}$ as a function of the free parameter $\rho_0$ appearing in equation (\ref{FORCE}).  Collecting the different terms coming from (\ref{SURFACEFINAL}), we obtain the following
\begin{eqnarray}
\label{sigmaAB}
\sigma_{AB} = c_S^4 \int_{flat} dx \left[ -\frac{G_{A2}}{2} (\partial_x \Psi_A)^2 -  \frac{G_{B2}}{2} (\partial_x \Psi_B)^2 - \frac{G_{AB}}{\rho_0^2} \partial_x \rho_A \partial_x \rho_B  + \tau^2 \theta^2(\tau) \frac{\rho_A\rho_B}{\rho} \left( \frac{\partial_x\rho_A}{\rho_A}-\frac{\partial_x \rho_B}{\rho_B} \right)^2\right].
\end{eqnarray}
The exact computation of the integral in equation (\ref{sigmaAB}) requires the  knowledge of the functions $\rho_A(x)$ and $\rho_B(x)$. However, useful insight can be gained by assuming that the sum of the two densities, $\rho_A + \rho_B = \langle \rho \rangle $ is constant and that the leading contribution to the integral comes from the interface  region, where $\rho_A \approx \rho_B$.  We can then  expand about the point $\phi = \rho_A - \rho_B = 0$ that we consider located at the central point $x_c$.  With these assumptions, we write 
$$\rho_A \approx \rho_B = \frac{\langle \rho \rangle}{2} \hspace{.2in} \partial_x \rho_A = -\partial_x \rho_B \approx \partial_x \phi  |_{x_c} , $$
$$\partial_x \psi_{A} \approx e^{-\langle \rho \rangle/2 \rho_0} \; \partial_x \rho_{A} |_{x_c} \hspace{.2in} \partial_x \psi_{B} \approx e^{-\langle \rho \rangle/2 \rho_0} \; \partial_x \rho_{B} |_{x_c}. $$ 
In this way, for $\tau=1$ (LBU), equation (\ref{sigmaAB}) finally delivers
\begin{equation}
\label{surften}
\sigma_{AB} \approx  \frac{\delta_w}{2} \Sigma_{AB}(\rho_0)(\partial_y \phi|_{x_c})^2
\end{equation}
with $\Sigma_{AB}(\rho_0)$ depending on the couplings and the parameter $\rho_0$ as follows:
\be\label{surften}
\Sigma_{AB}(\rho_0)=\left( \frac{2 G_{AB}}{\rho_0^2} + \frac{2}{\langle \rho \rangle} - (G_{A2}+G_{B2}) e^{-\langle \rho \rangle/\rho_0}  \right).
\ee
and where $\delta_w$ is the characteristic thickness of the interface.  Equation (\ref{surften}) shows that by increasing $\rho_0$ the surface tension can be made negative, so that the condition $\Sigma_{AB}(\rho_0)=0$ stipulates a vanishing surface tension. Indeed, upon increasing $\rho_0$, the positive contribution of repulsive interactions is  weakened, whereas the negative contribution of self-interactions is enhanced, provided that $G_{A2}$ and $G_{B2}$ are both positive. In figure \ref{fig:6} we show the analytical computation of $\Sigma_{AB}$ as  a function of $\rho_0$ for the set of parameters $\langle \rho \rangle=1.23$, $G_A^a = -15$, $G_A^r = 14.1$, $G_B^a = -14$, $G_B^r = 13.1$, $G_{AB} = 0.405$, corresponding to $G_{A2} = 9.17$ and $G_{B2} = 8.46$, all in LBU.  The theory predicts a crossover of the surface tension to negative values at $\rho_0 \sim 0.72$, quite close to the numerically observed result $\rho_0 \sim 0.717$ (see figure \ref{fig:7}). This shows that the interplay between inter-species repulsion and intra-species repulsion/attraction is key to attain vanishing small values of the surface tension, which are in turn crucial to reproduce the physical properties described in the second part of this paper.

\section{Numerical Results} 

Having discussed the major theoretical aspects of this model, we next proceed to present  the results of numerical simulations.  The baseline simulations are performed on a $2$ dimensional grid  $N_x \times N_y =128 \times 128$, with occasional 
enlargements to $N_x \times N_y =256 \times 256$ and $N_x \times N_y =512 \times 512$.  The two fluids are initialized with zero speed and random initial conditions for the two densities $\rho_A$ and $\rho_B$.  More specifically, we choose $\lra{\rho_A}=\lra{\rho_B}=0.612$, with a standard deviation  $\pm 0.01$ from the background density value.  The couplings have been set to the following values in LBU:
\begin{equation}\label{STANDARD07}
\begin{cases}
G_A^a = -15.0, \hspace{.2in} G_A^r = 14.1\\
G_B^a = -14.0, \hspace{.2in} G_B^r = 13.1\\
G_{AB} = 0.405 \\
\end{cases}
\end{equation}
defined as standard set at $\rho_0=0.7$ and
\begin{equation}\label{STANDARD083}
\begin{cases}
G_A^a = -9.0, \hspace{.2in} G_A^r = 8.1 \\
G_B^a = -8.0,  \hspace{.2in} G_B^r = 7.1 \\
G_{AB} = 0.405
\end{cases}
\end{equation}
defined as standard set at $\rho_0=0.83$. The relaxation time is fixed to $\tau=1$ (LBU), corresponding to a kinematic viscosity $\nu=1/6$ (LBU). The corresponding value of the surface tension is approximately $\sigma_{AB} \sim 0.01$ in both standard sets. The main difference between the two sets of parameter is that the standard set at $\rho_0=0.83$ displays a more refined (in terms of computational grid points) interface. Moreover, the standard sets of parameters have been chosen in such a way that both components $A$ and $B$ are in the dense (liquid) phase.

\section{Free dynamics of the density configuration}

\begin{figure}
\includegraphics[scale=0.5]{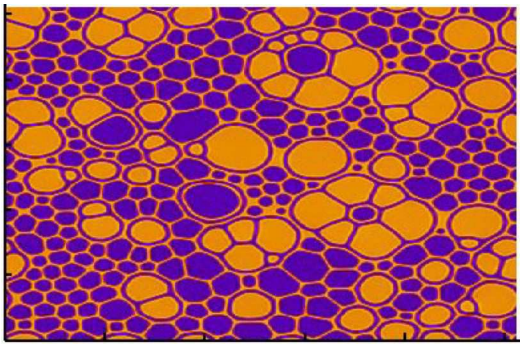}
\caption{Contours of the density field $\rho_A(x,y)$  obtained with a numerical simulation on  a $N_x \times N_y= 512 \times 512$ grid. 
The parameters are those of the standard set at $\rho_0=0.83$, as given in equations (\ref{STANDARD083}).}\label{fig:8}
\end{figure}

We begin by investigating the free configurational dynamics  of the density field under the sole effect of internal interactions (no-forcing).  The first observation is that, even after a very long time-span (hundreds of thousands time-steps) the fluid densities $\rho_A(x,y)$ and $\rho_{B}(x,y)$ do not exhibit any macroscopic separation between the two fluids A and B.  Instead, a multitude of metastable domains ("droplets") of fluid A in fluid B and viceversa is observed, as a result of the complex interplay between repulsive  (short-range inter-species and mid-range intra-species) and attractive  (short-range intra-species) interactions. This is in  line with other studies in solid state physics and soft matter \cite{Seul95,Stojkovic99,Reichhardt03,Tuzel07}. The final result is a rich configurational structure of the density field, as   shown in figure \ref{fig:8}. The most salient feature of the density configurations is the formation of 'belts' of fluid A (B), entrapping bubbles of both fluids B and A inside. As we shall see shortly, these belts exert a major influence on the rheology of the fluid, and in particular, their formation/rupture is responsible for a number of features, such as dynamical heterogeneity and arrest, long-time  relaxation, ageing effects and intermittency.

\begin{figure}
\includegraphics[scale=1]{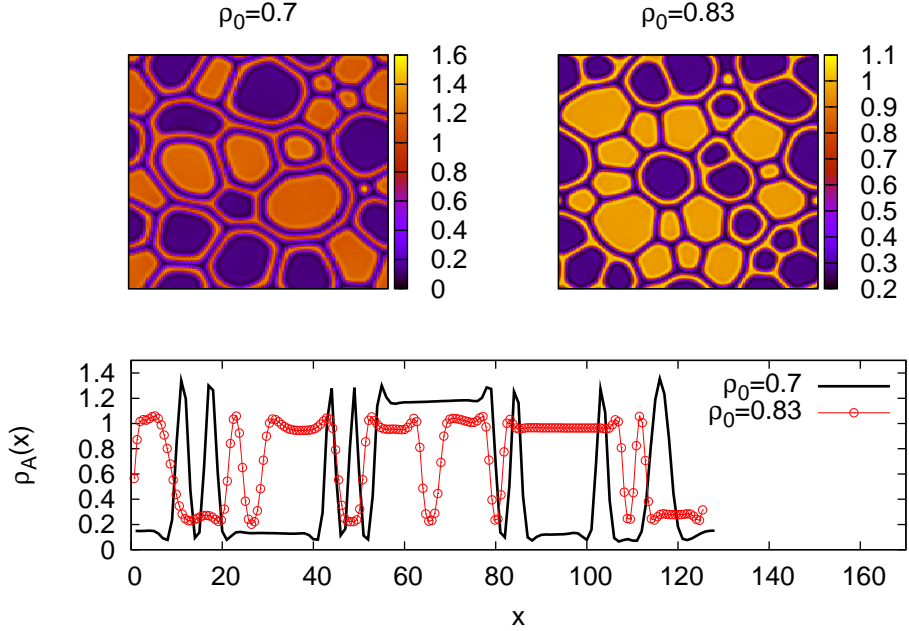}
\caption{Contours (top) and centerline cuts (bottom) of the density of  the specie A, for the standard set of parameters at $\rho_0=0.7$ (see equations set (\ref{STANDARD07})) and at $\rho_0=0.83$ (see equations set (\ref{STANDARD083})). All results are reported in LBU.}
\label{fig:9}
\end{figure}

The occurrence of belts of fluid A (B) entrapping fluid B (A), is well visible in figure \ref{fig:9}, where also shown (bottom panel) are the density cuts of species A, across the midline $y/N_y=0.5$ for the two different standard sets of parameters at $\rho_0=0.7$ (see equations set (\ref{STANDARD07})) and $\rho_0=0.83$ (see equations set (\ref{STANDARD083})) . Although the details of the density contours and profiles are clearly different in the two cases, the main qualitative feature, namely the presence  of a multitude of metastable "droplets" of both fluids A and B, is well visible in both cases. Therefore, these "droplets" are naturally interpreted as the metastable structures  which permit the two-fluid system to escape the fully-separated  minimum-interface configuration.  

\section{Dynamic response under applied shear}

In view of the rich morphology of the density field discussed in the previous section, it is natural to inspect the behavior of the two-fluid system under the effect of an external drive. To this purpose, we analyze the dynamic response to an externally applied  shear flow of the form  $U_x(x,y) = U_0 \sin (ky)$, $U_y=0$, with $k=1$.  This is realized by imposing a volumetric body force in the LB equation. The rheological properties of the fluid are measured by monitoring the following response function: 
\begin{equation}\label{RESPONSE}
R(t)=\frac{\hat{\bar U}(k=1;t)}{U_0} \equiv \frac{\nu_0}{\bar{\nu}(t)}
\end{equation}
where $\hat{\bar U}(k;t)$ is the Fourier transform of the line-averaged speed along 
the $x$ direction, $\bar{U}(y;t)=\sum_{x} U(x,y;t)/N_x$,  $\nu_0$ is the
nominal kinematic viscosity of both fluids and $\bar{\nu}$ defines the effective 
viscosity of the two-fluid system.  By construction, under undisturbed flow conditions,  $R>0$, so that $R\ll 1$ provides a  direct measure of slowing-down through enhanced effective viscosity and  eventually, structural arrest ($R=0$).  Baseline simulations are performed on a $N_x \times N_y = 128 \times 128$ grid, for up to $5\times 10^6$ LBU time steps.

\subsection{Cage formation and rupture}

\begin{figure}
\includegraphics[scale=0.8]{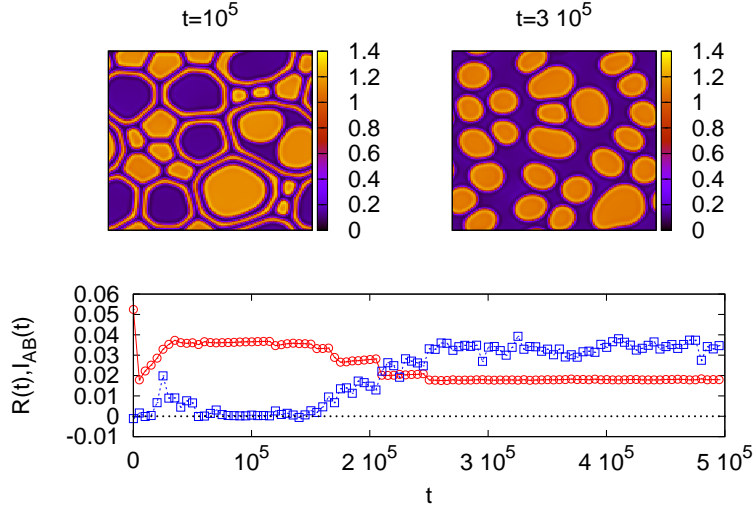}
\caption{\label{fig:10}
Response function $R(t)$ (squares)  given 
in equation (\ref{RESPONSE}) and surface indicator $I_{AB}(t)$ (circles) 
given in equation (\ref{INDICATOR}) for the standard set of parameters at $\rho_0=0.7$ (see equations set (\ref{STANDARD07})).  The upper panels show two snapshots of the density field in a blocked and flowing state, respectively.  
Note that the flowing state is nonetheless characterized by a small fraction (a few percent) of the undisturbed flow speed, corresponding to values or $R$ larger than $0$. The dotted line at zero is reported for visual guidance.  All results are given in LBU.}
\end{figure}

A typical response function is shown in figure \ref{fig:10} (lower panel), together with two snapshots of the density contours at $t=10^5$ and $t=3 \; 10^5$ (upper panel).  In the same figure, also shown is an indicator of the interface  area (length in 2d) between the two fluids, defined as follows: 
\be\label{INDICATOR}
I_{AB}(t) = - \sum_{x,y} {\bm \nabla} \rho_A(x,y;t) \cdot {\bm \nabla} \rho_B (x,y;t).
\ee
This figure provides a neat example of dynamical arrest (between $t=5 \hspace{.1in} 10^4$ LBU and $t \sim 1.5 \hspace{.1in} 10^5$ LBU, followed 
by a progressive recovery of the flow  (from $t \sim 1.5 \hspace{.1in} 10^5 $ LBU to $t \sim 3 \hspace{.1in} 10^5$ LBU, until 
the system starts to flow again, although with a 25-fold higher viscosity than the nominal one, i.e. $R \sim 0.04$ versus $R=1$. 

The two snapshots refer to a blocked configuration ($t=10^5$ LBU) and to a flowing one ($t=3 \hspace{.1in} 10^5$ LBU), respectively.  In the former, belts caging one fluid into another are well visible, which subsequently break  down and disappear, thereby allowing the system to flow again. Consistently with this picture of cage rupture and annihilation, the interface length, as measured by $I_{AB}$, is seen to decrease in going from the arrested to the cage-free flowing configuration.  This picture clearly illustrates the vital role played by the cage structures  on the global rheology of the two-fluid system. It is worth emphasizing that, due to the mesoscopic nature of the present model, the rupture  of a single cage, corresponds to a large collection of atomistic events, and consequently it leads to observable effects on the overall rheology of the system.  

Next, we investigate the time dependence of the response function for different values of the shear forcing $U_0$.

\begin{figure}
\includegraphics[height=6cm,width=8cm]{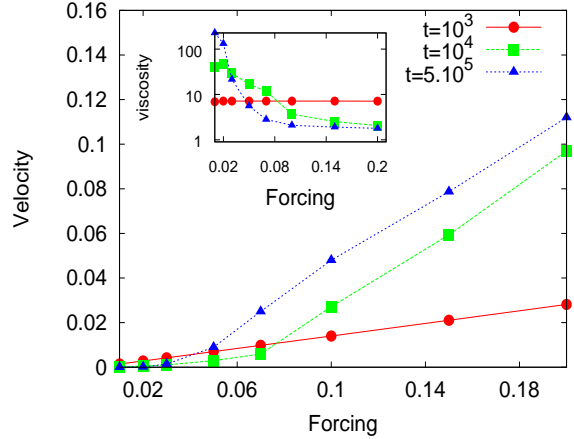}
\caption{\label{fig:11} The response function $R(t)$ at three-different instants, $t=10^3, 10^4, 5 \;10^5$ LBU, as a function of the applied forcing $U_0$. Simulations are carried out for the standard set of parameters at $\rho_0=0.7$ (see equations set (\ref{STANDARD07})). The inset reports the effective viscosity, i.e. the system average velocity versus  the forcing amplitude. All results are reported in LBU.} 
\end{figure}

In figure \ref{fig:11}, we show a typical example for the response function $R(t)$ at three-different  instants, $t=10^3,10^4,5 \;10^5$ LBU, and for different values of the forcing $U_0$. At short times the response is linear with $U_0$ for all investigated values of $U_0$,  (Newtonian behavior). At longer times, however, a typical yield-stress threshold appears, i.e. the fluid starts to flow only beyond a critical value of the forcing, $U_0 \sim 0.03$ LBU.  Above this threshold, the fluid starts to flow at a higher rate (see also inset, reporting the effective viscosity) as compared to the short-time response, thereby providing evidence of non-newtonian, shear-thinning, behavior.

\subsection{Dynamics of correlations: ageing effects}

\begin{figure}[h]
\includegraphics[scale=0.75]{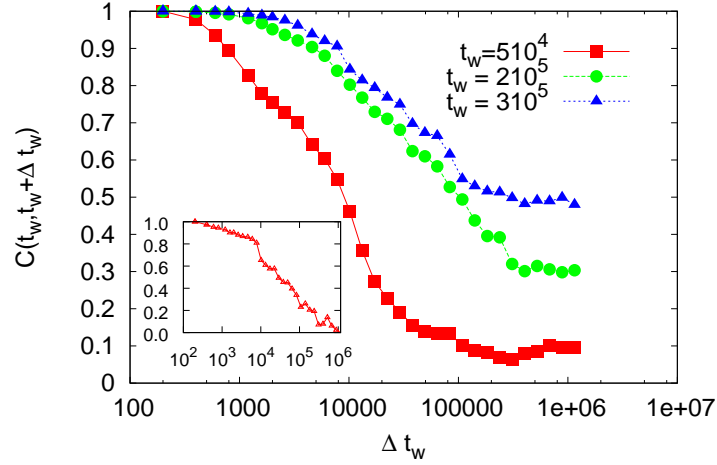}
\caption{\label{fig:12}Ageing of the system.  Correlation function as defined in (\ref{CORRE}) computed for different waiting times $t_w$  ($t_w = 5~10^4 $, red squares,  $t_w = 2~10^5$,  green circles and $t_w=3 \; 10^5$, blue triangles ) with  shear stress $U_0 = 0.02$ LBU. The waiting time $\Delta t_w$ is reported on a log scale. Simulations are carried out for the standard set of parameters at $\rho_0=0.7$ (see equations set (\ref{STANDARD07})). The inset reports the correlation function for $t_w=3~10^5$ and $U_0=0.03$: with  increasing shear stress the structural arrest disappears, as witnessed by  a vanishing value of the correlation function in the limit $\Delta t_w \rightarrow \infty$. All results are reported in LBU.}
\end{figure}

\begin{figure}[h]
\includegraphics[scale=0.75]{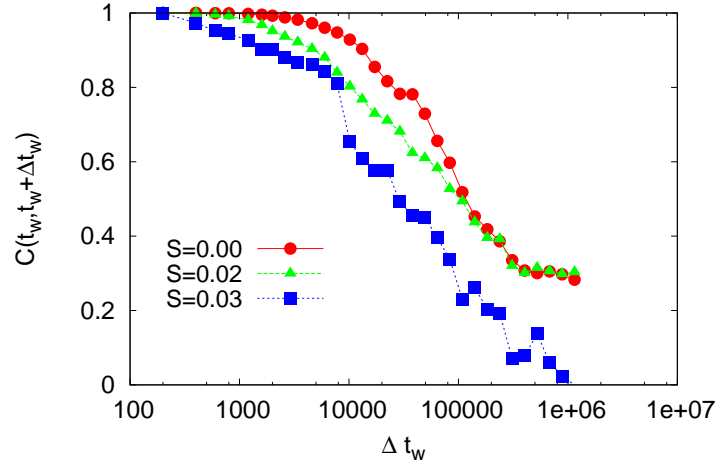}
\caption{\label{fig:13} 
Ageing of the system at increasing shear stress.  This configuration of the system represents a threshold-yield fluid.   The system does not flow and shows ageing until a certain threshold of the applied  forcing, above  which the system decorrelates completely.  In the figure $S$ is the externally imposed shear and the waiting time $\Delta t_w$ is reported on a log scale. All results are reported in LBU and the parameters are those of the standard set at $\rho_0=0.7$ (see equations set (\ref{STANDARD07})). }
\end{figure}
We next inspect another typical phenomenon of soft-glassy matter, namely ageing. 
To this purpose, following upon the spin-glass literature \cite{CAV08}, we define  the order parameter $\phi \equiv \rho_A-\rho_B$ and compute its {\it overlap}, defined  through the autocorrelation function:
\be
\label{CORRE}
C(t_w,\Delta t_w) = \frac{\langle \sum_{x,y} \phi(x,y;t_w) \phi(x,y;t_w+\Delta t_w) \rangle}
{\langle \sum_{x,y} \phi(x,y;t_w) \phi(x,y;t_w) \rangle}
\ee
where $t_w$ is the waiting time, $\Delta t_w$ is the time lapse between the two
density configurations and brackets stand for averaging over an ensemble of realizations. In figure \ref{fig:12}, we show the correlation function corresponding to  three different waiting times, $t_w$ ($t_w = 5 \;10^4$ LBU  , red squares,   $t_w = 2 \; 10^5$ LBU,  green circles and $t_w=3 \; 10^5$ LBU, blue triangles), for a forcing amplitude $U_0 = 0.02$ LBU. Ageing effects are clearly visible, in the form of a dependence of the time-decay of the correlation function on the waiting time $t_w$, and, more specifically, with an increasingly slower decay as the waiting time is increased. Moreover, the correlation function saturates to a non-zero value in the long-time limit  (broken ergodicity), which is another typical signature of structural arrest (the system does not succeed to fully decorrelate).  This behavior shows qualitative changes upon increasing the forcing term. In the inset of the same figure, we show the correlation function for $t_w=3 \; 10^5$ LBU and  a slightly larger forcing, $U_0=0.03$ LBU.  With increasing shear stress, cages are broken, and the structural arrest disappears, thereby allowing the correlation function to decay to zero (see figure \ref{fig:13}).   The disappearance of structural arrest under sufficiently strong shear  is again a distinctive feature of flowing soft-glassy materials \cite{Coussot02} and these results are in qualitative agreement with molecular dynamics simulations \cite{Barrat02}.

\subsection{Intermittency and Barkhausen noise}

\begin{figure}
\includegraphics[scale=0.75]{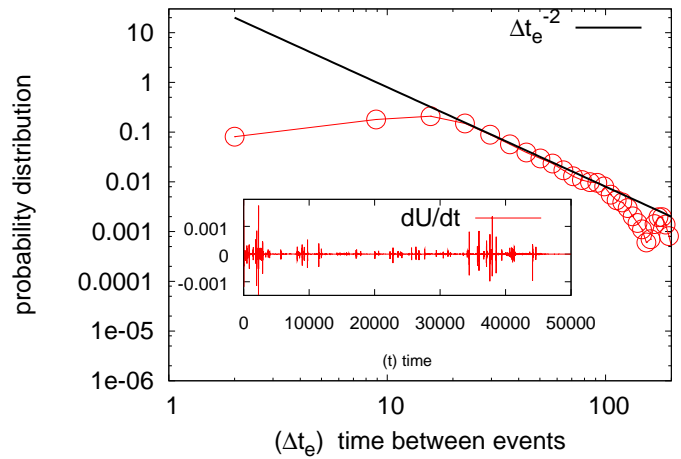}
\caption{\label{fig:14} Time evolution of the time-derivative of the average fluid velocity $U(t)$ (inset), showing a characteristic intermittent behavior. In the main picture, shown is the probability distribution of the time-lapse ($\Delta t_e$) between two subsequent bursts of $dU/dt$. A typically power law decay $\approx \Delta t_e^{-\alpha}$, with exponent $\alpha=2$ is observed. The figures refer to the standard set of parameters with $\rho_0=0.7$ as given in equations set (\ref{STANDARD07}) and results are all expressed in LBU.}
\end{figure}
Barkhausen noise is a well-known phenomenon displayed by disordered ferromagnetic samples under the effect of a slowly-changing magnetic field \cite{BARK}. A small ramp-up in the magnetic field triggers one domain and the perturbation spreads to neighboring domains, producing an avalanche which results in a series of jumps in the magnetization, as the systems transits from one metastable state to another. Several experiments show that the distribution of size, duration and energy of the Barkhausen jumps exhibit a power-law decay. The present two-fluid model also shows evidence of Barkhausen-like  intermittency in the time-derivative of the response function. In figure \ref{fig:14}, we show the probability distribution of the time-lapse $\Delta t_e$ between subsequent bursts (also called 'events')  of the response function (see inset). Interestingly, such distribution follows a power-law distribution  $\sim \Delta t_e^{-\alpha}$, with $\alpha \sim 2$. This invites a further analogy between the fluid cages discussed previously and the magnetic domains responsible for Barkhausen effects in disordered ferromagnets. The systematic exploration of the dependence of  these Barkhausen-like effects on the various parameters of our system, is left as an interesting topic for future research.

\section{Sensitivity to initial conditions}

\begin{figure}
\includegraphics[scale=0.75]{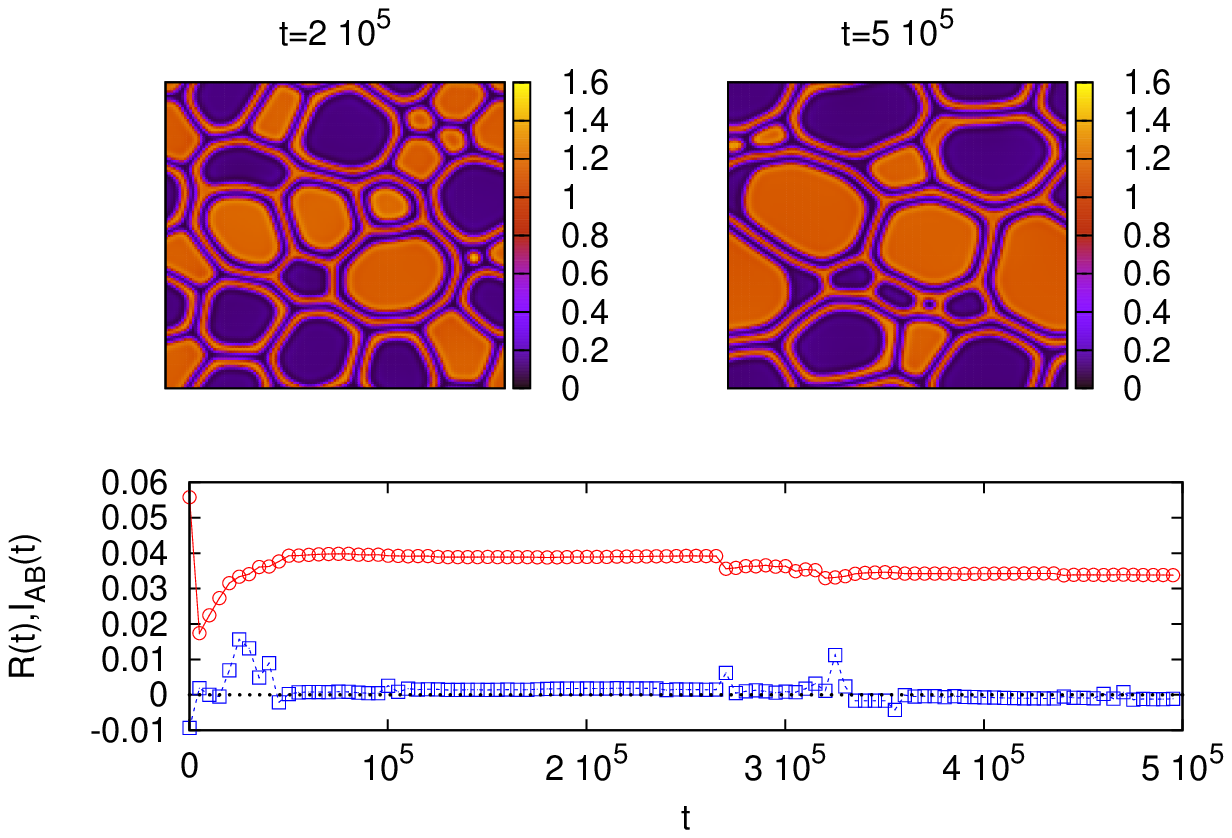}
\includegraphics[scale=0.75]{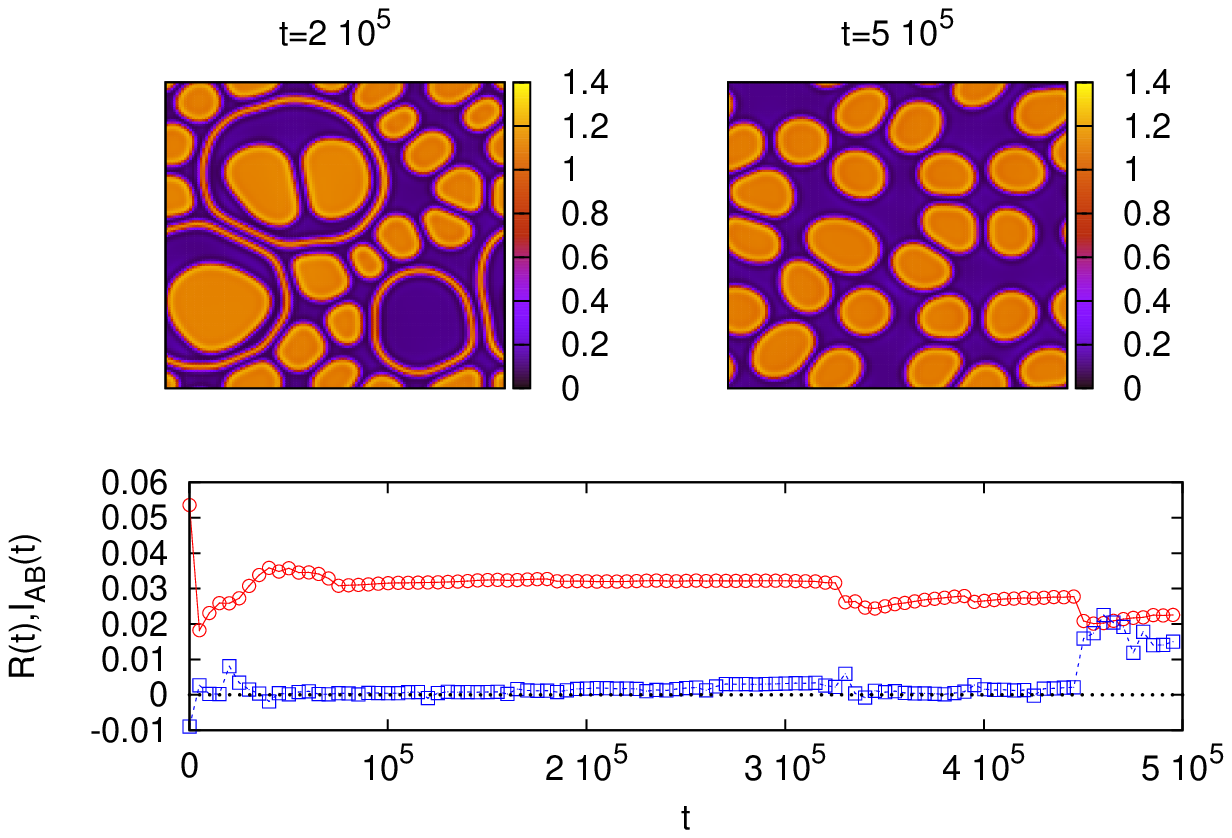}
\caption{\label{fig:15} 
Response function $R(t)$ (squares) given in equation (\ref{RESPONSE}) and 
surface indicator $I_{AB}(t)$ (circles) given in equation (\ref{INDICATOR}) for 
two different realizations of the initial conditions, obtained by changing the seed of the random number generator.  The runs are performed with the standard set of parameters at $\rho_0=0.7$ as given in equations set (\ref{STANDARD07}).  The dotted line at zero is reported as a visual guidance. All results are reported in LBU.}
\end{figure}

\begin{figure}
\label{fig:17}
\includegraphics[scale=0.8]{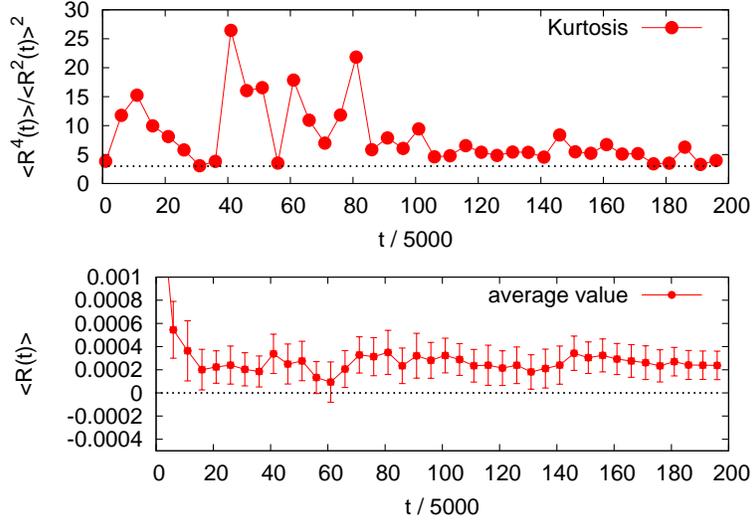}
\caption{\label{fig:16} Evaluation of the response function $R(t)$ under the application of a  shear forcing $U_0$ kept fixed in the numerical simulations. We have chosen $200$ equispaced times between $t=0$ and $t=10^6$ (LBU) and run $100$ numerical simulations by changing the random initial conditions.  The upper plot shows the Kurtosis, ${\cal K}(t)=\frac{\langle R(t)^4 \rangle}{\langle R(t)^2 \rangle^2}$, where $\langle ... \rangle$ refers to the average over the various numerical runs at a fixed $t$.  Intermittency is clearly visible from this plot. For the sake of clarity, we have reported the value of the Kurtosis for a Gaussian variable (equal to $3$, dotted line).  In the lower plot, the response function and the variance (errorbars) is then evaluated and plotted as a function of time. The line at zero is reported as a visual guideline. For the numerical simulations, we have used the standard set of parameters at $\rho_0=0.7$ as given in equations set (\ref{STANDARD07}), with a shear forcing term $U_0/N_x = 0.1/128$ in LBU. The resolution is $N_x \times N_y= 128 \times 128$.} 
\end{figure}

In the previous sections we have illustrated several phenomena typical of soft-glassy materials. A natural question arises as to the degree of robustness of these phenomena towards  changes in the initial conditions and size of the system.  Although a systematic exploration of these effects requires a study of its own, in the following we provide some preliminary information. As expected, the detailed dynamics of the response function shows a strong sensitivity to the noise realization, with some configurations reaching a plateau in the early stage of the evolution (see figure \ref{fig:15}), while others never attaining any plateau within the entire simulation span. In order to probe the robustness of the response function $R(t)$ towards changes in  the random realization of the initial conditions, we have performed a series of  $100$ simulations by changing the noise realization at a fixed variance of the initial density.   Notwithstanding the qualitative differences in the detailed response function, the main picture portrayed in the previous sections, namely arrested flow due to formation of fluid cages, and restored flow upon cage rupture, is found to apply to all simulations. To better appreciate the statistical dynamics of the present system, in figure \ref{fig:16}, top panel, we  show the time evolution of the Kurtosis ${\cal K}(t)=\frac{\langle R(t) ^4\rangle}{\langle R(t) ^2\rangle ^2}$ of the response $R(t)$, as computed  from the set of $100$ realizations. This figure shows clear evidence of large fluctuations in the first half of the evolution, followed by a more quiescent stage in the second half. To be noted that, even in the quiescent stage, the Kurtosis is still around ${\cal K} (t) \sim 5$, hence well above the Gaussian value ${\cal K} (t)=3$, thereby confirming the strongly fluctuating nature of the phenomenon. A similar message is conveyed by the bottom panel of the same figure, which reports the average value $\langle R(t) \rangle$, along with the variance, as a function of time. From this figure, we see that the variance is generally comparable to the mean value, sometimes even larger. The intermittent nature of the response $R(t)$ is further highlighted in figure \ref{fig:17}, which shows the probability distribution function of $R(t)$, sampled over three close-by time-slices. This pdf exhibits intermittent tails on both negative and positive sides, with a slight prevalence of the latter, consistently with the positive sign of $\langle R(t) \rangle$.

\begin{figure}
\label{fig:17}
\includegraphics[scale=0.75]{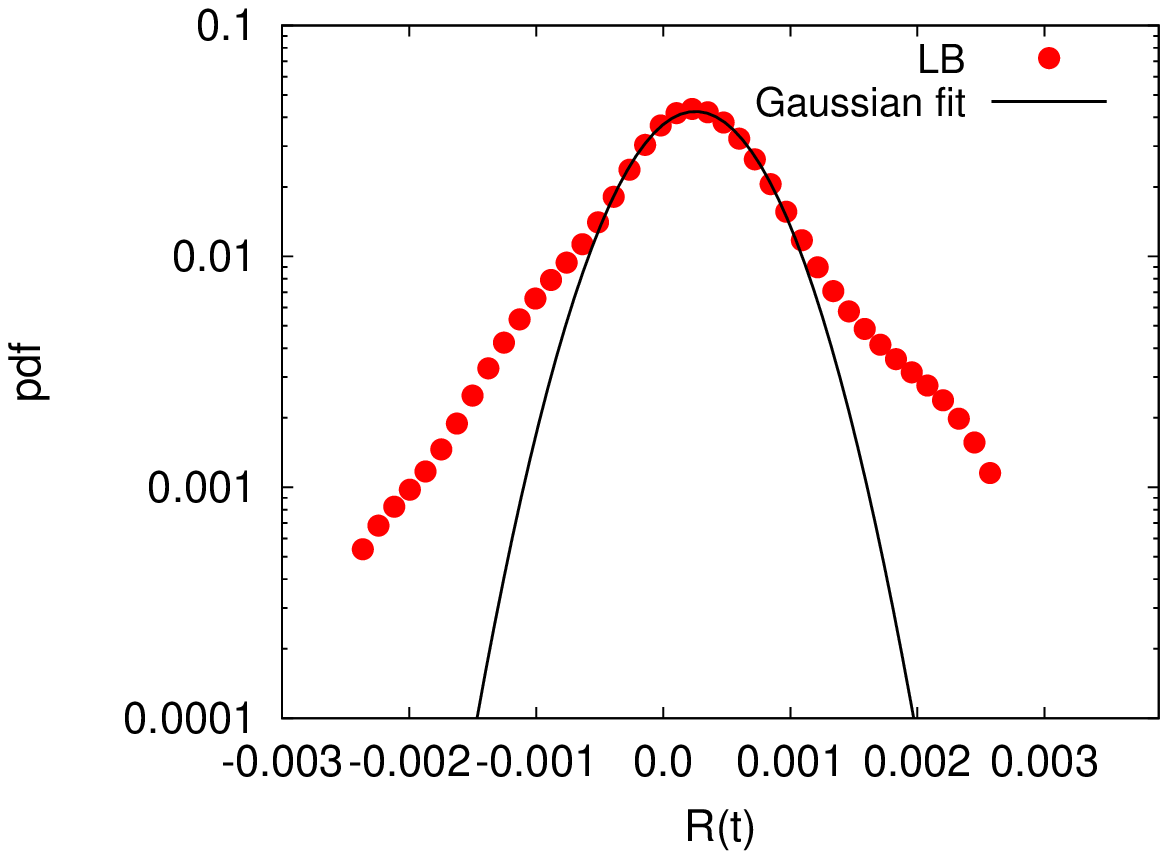}
\caption{\label{fig:17} Plot of the probability density function for the response function $R(t)$ under the application of a  shear forcing $U_0/N_x$, kept fixed in the numerical simulations. The probability has been obtained from the analysis of the response function at three distinct but very close times $t=4.8~10^5, 4.9~10^5, 5~10^5$ (LBU) in $100$ numerical simulations with different random initial conditions (the choice of $3$ close instants is meant to enhance the statistics). The numerical simulations are performed with the standard set of parameters at $\rho_0=0.7,$ as given in equations set (\ref{STANDARD07}) with a shear forcing term $U_0/N_x = 0.1/128$ in LBU.  The resolution is $N_x \times N_y= 128 \times 128$. } 
\end{figure}

\section{Summary and outlook}

Summarizing, we have provided a theoretical analysis of a two-component lattice Boltzmann model with mid-range intra-molecular repulsion  and short-range inter-molecular repulsion.  In particular, equilibrium densities and the surface tension as a function of the main parameters of the model, have been computed and shown to exhibit satisfactory agreement with numerical tests. We have also presented a series of numerical simulations proving the capability of this system of reproducing many distinctive features of soft material  behavior, such as slow-relaxation, anomalous enhanced viscosity, caging effects, aging  under shear and Barkhausen intermittency. The present lattice kinetic model caters for this very rich physical picture  at a computational cost only marginally exceeding the one for a simple fluid. As a result, it should be possible to use it for future investigations of the non-equilibrium rheology. In particular,  it may be  useful to  get new insights  in  the  coexistence  of  liquid and  solid  regions  (shear localization,  shear  banding,  cracks)  as  observed  with  emulsions \cite{Coussot02,Becu06},   foams    \cite{Debregeas01,Kabla03,Janiaud06},   worm-like   micelles \cite{Salmon03,Lerouge06}  and granular  materials \cite{Huang05,Mills08}. Still,  such a hydro-kinetic  method  might be  interesting  to  treat  the issue  of dilatancy in foams observed in recent experiments \cite{Marze05}.  In order to analyze those systems, on going research is devoted to a systematic investigation of the  system behavior at different concentrations of  the  two  species,  its  sensitivity  to  initial  conditions  and finite-size  effects,  as  well  as  its  response  to  time-dependent loads.  Also of  current  interest are  extensions to  three-component fluids, in order  to account for the explicit  presence of surfactants \cite{BOGOS}.

\section*{Appendix: heuristic mapping to physical units}
One of main advantages of the present mesoscopic approach 
is to provide access to hydrodynamic scales at an affordable computational cost.
In order to appreciate this point, it is of interest to discuss the
conversion between LB and physical units.
The spatial units, namely the LB spacing $\Delta x$, can be estimated 
by fixing the surface tension according to the following relation (subscript $phys$ denotes physical units):
\begin{equation}
\sigma_{phys} \approx \sigma_{LB} \frac {kT}{(\Delta x)^2}
\end{equation}
where the subscript LB denotes the value in LBU. For micro-emulsions, we may estimate $\sigma_{phys} \sim 10^{-4}$ $N/m$, so that at standard conditions ($T=300^{\circ}$), a LB surface tension $\sigma_{LB} \sim 0.01$ corresponds to $\Delta x = 
\sqrt{kT \frac{\sigma_{LB}}{\sigma_{phys}}} \sim \;10^{-9}$ $m$. This means that a $N_x \times N_y = 128 \times 128$ simulation covers a squarelet of  about $0.1$ micron in side.     Similarly, the time units (the LB time step $\Delta t$)  can be estimated by fixing the kinematic viscosity according  to the relation:
\begin{equation}
\nu_{phys} \approx \nu_{LB} \frac {(\Delta x)^2}{\Delta t}
\end{equation}
By taking $\nu_{phys} \sim 10^{-6}$ $m^2/s$ and $\nu_{LB} \sim 0.1$, a lattice
spacing $\Delta x \sim 10^{-9}$ $m$, would yield $\Delta t \sim 10^{-13}$ s.
As a result, a $10^6$ time-step simulation covers about $0.1$ $\mu s$.
These values are only marginally higher than those 
typically used in Molecular Dynamics simulations.
However, the point is that the present model lends itself to substantial
upscaling both in space and time, while still presenting an affordable 
computational cost. For instance, preliminary simulations on a $N_x \times N_y = 1024 \times 1024$ grid, span $10^6$ lattice time steps in about one-day elapsed time on Graphical Processing Units architecture \cite{Bernaschi09}. Such simulations cover a square domain about some  microns in side, over a time span of some microseconds, a way beyond the capabilities of standard Molecular Dynamics or Monte Carlo simulations.

\section{Acknowledgments}

SS wishes to acknowledge financial support from the project INFLUS (NMP3-CT-2006-031980) and  SC financial support from the ERG EU grant and consorzio COMETA.
Fruitful discussions with J.-F. Berret, L. Biferale, M. Cates, A. Cavagna, C. Gay, D. Nelson, G. Parisi, N. Rivier, S. Lerouge, and F. Toschi are kindly acknowledged.


\begin{thebibliography}{99}

\bibitem{Larson99} R.G. Larson, {\it The Structure and Rheology of Complex Fluids} (New York, Oxford university press, 1999)

\bibitem{Coussot05} P. Coussot, {\it Rheometry of pastes, suspensions, and granular materials} (Wiley-Interscience, 2005)

\bibitem{Chaikin95} P.M. Chaikin \& T.C. Lubensky, {\it Principles of Condensed Matter Physics} (Cambridge University Press, Cambridge, 1995)

\bibitem{Lyklema91} J. Lyklema, {\it Fundamentals of Interface and Colloid Science} (Academic Press, London, 1991) 

\bibitem{Evans99} D. F. Evans \& H. Wennerstrm, {\it The Colloidal Domain} (Wiley-VCH, New York, 2nd edition, 1999)

\bibitem{Degennes79} P.G. De Gennes, {\it Scaling Concepts in Polymer Physics}  (Cornell University Press, Ithaca, 1979)

\bibitem{Doi86} M. Doi \& S. F. Edwards, {\it The Theory of Polymer Dynamics} (Oxford University Press, Oxford, 1986) 


\bibitem{Grosberg94}  A.Y. Grosberg \& A. R. Khokhlov, {\it Statistical Physics of Macromolecules} (AIP Press, New York, 1994) 


\bibitem{Weaire99} D. Weaire \& S. Hutzler, {\it The Physics of Foams} (Oxford  University Press, 1999). 

\bibitem{Russel89} W.B. Russel, D.A. Saville \& W.R. Schowalter, {\it Colloidal Dispersion} (Cambridge University Press, Cambridge England, 1989) 

\bibitem{Poole92} P.H. Poole, F. Sciortino, U. Essmann \& H. E. Stanley, {\it Nature} {\bf 360}, 324 (1992)

\bibitem{Sollich97} P. Sollich, F. Lequeux, P. H\'ebraud \& M. E. Cates, {\it Phys. Rev. Lett.} {\bf 78}, 2020 (1997) 


\bibitem{Eckert02} T. Eckert \& E. Bartsh, {\it Phys. Rev. Lett.} {\bf 89}, 125701 (2002) 
        
\bibitem{Sciortino02} F. Sciortino, {\it Nat. Mat.} {\bf 1}, 145 (2002) 

\bibitem{Pham04} K.N. Pham, A.M. Puertas, J. Bergenholtz, S.U. Egelhaaf, A. Moussaf¨id, P.N. Pusey, A.B. Schofield, M.E. Cates, M. Fuchs \& W.C.K. Poon, {\it Science} {\bf 296}, 104 (2004)

\bibitem{Guo07} H. Guo, J. N. Wilking, D. Liang, T. G. Mason, J. L. Harden \& R. L. Leheny, {\it Phys. Rev E} {\bf 75}, 041401 (2007)


\bibitem{Schall07} P. Schall, D. A. Weitz \& F. Spaepen, {\it Science} {\bf 318}, 1895 (2007) 


\bibitem{Lu08} P. J. Lu, E. Zaccarelli, F. Ciulla, A. B. Schofield, F. Sciortino \& D. A. Weitz, {\it Nature} {\bf 453}, 499 (2008) 

\bibitem{Allen90}  M.P. Allen \& D.J. Tildesley, {\it Computer simulations of liquids} (Oxford University Press, New York, 1989) 

\bibitem{Frankel96} D. Frankel, \& B. Smith, {\it Understanding molecular simulation} (Academic Press, San Diego, 1996) 

\bibitem{Binder97} K. Binder \& D.W. Herrman, {\it Monte Carlo simulation in Statistical Physics} (Springer, Berlin, 1992)


\bibitem{ALE74}C.W. Hirt, A.A. Amsden \& J.L. Cook, {\it J. Comp. Phys.} {\bf 14} 227–253 (1974). 

\bibitem{VOF99} R. Scardovelli \& S. Zaleski, {\it Annu. Rev. Fluid Mech.} {\bf 31}, 567 (1999) 

\bibitem{Quarteroni07} C. Canuto, M. Y. Hussaini, A. Quarteroni \& T. A. Zang, {\it Spectral Methods: Evolution to Complex Geometries and Applications to Fluid Dynamics}, (Springer, Berlin, 2007)


\bibitem{Kob02} W. Kob in {\it Slow relaxation and nonequilibrium dynamics in condensed matter}, Les Houches, Session LXXVII, J.-L. Barrat, M. Feigelman, J. Kurchan \& J. Delibard edts., Springer-EDP sciences (2002)

\bibitem{Hoogerbrugg92} P. J. Hoogerbrugge \& J. M. V. A. Koelman, {\it Europhys. Lett.}, {\bf 19}(3), 155  (1992)

\bibitem{Cun79} P. A. Cundall \& O. D. L. Strack, {\it Geotech.} {\bf 29}, 47 (1979)

\bibitem{Dur87} L. Durlofsky, J. Brady \& G. Bossis, {\it J. Fluid Mech.} {\bf 180}, 21 (1987)

\bibitem{Duri95} D. J. Durian, {\it Phys. Rev. Lett.} {\bf 75}, 4780 (1995)

\bibitem{GDR04} Gdr Midi, {\it Euro. Phys. J. E} {\bf 14}, 341 (2004)

\bibitem{Bra01} J. Brady, {\it Chem. Eng. Sci.} {\bf 56}, 2921 (2001)

\bibitem{Hol05} R. Hohler \& S. Cohen-Addad, {\it J. Phys.: Cond. Mat.} {\bf 17}, R1041 (2005)

\bibitem{DOI} M. Doi \& T. Ohta, {\it J. Chem. Phys.}  {\bf 95}, 1242-1248 (1991)


\bibitem{Gay08}P. Rognon, \& C. Gay, {\it Eur. Phys. J. E} {\bf 27}, 253-260 (2008)

\bibitem{Mcnamara98} G. R. McNamara \& G. Zanetti, {\it Phys. Rev. Lett.} {\bf 61}, 2332 (1988)  

\bibitem{Higuera89a} F. Higuera \& J. Jimenez, {\it Europhys. Lett.} {\bf 9}, 663 (1989) 

\bibitem{Higuera89b} F. Higuera, S. Succi \& R. Benzi, {\it Europhys. Lett.} {\bf 9}, 345 (1989)

\bibitem{Benzi92}R. Benzi, S. Succi \& M. Vergassola, {\it Phys. Rep.} {\bf 222}, 145, (1992)

\bibitem{BGK54} P.-L. Bathnagar, E. Gross \& M. Krook, {\it Phys. Rev.} {\bf 94}, 511-525 (1954)

\bibitem{Chen98} S. Chen \& G. Doolen, {\it Annu. Rev. Fluid Mech.} {\bf 30}, 329-364 (1998)

\bibitem{Gladrow00} D.A. Wolf-Gladrow, {\it Lattice-gas Cellular Automata and Lattice Boltzmann Models} (Springer, Berlin, 2000) 

\bibitem{SC_93} X. Shan \& H. Chen,  {\it Phys. Rev. E} {\bf 47}, 1815 (1993) 

\bibitem{SC_94} X. Shan \& H. Chen,  {\it Phys. Rev. E} {\bf 49}, 2941 (1994)

\bibitem{Benzi09} R. Benzi, S. Chibbaro \& S. Succi, {\it Phys. Rev. Lett.} {\bf 102}, 026002 (2009)

\bibitem{Sirovich62} L. Sirovich, {\it Phys. Fluids} {\bf 5}, 908 (1962) 

\bibitem{Hamel65} B.B. Hamel, {\it Phys. Fluids} {\bf 8}, 418 (1965) 

\bibitem{Hamel66} B.B. Hamel, {\it Phys. Fluids} {\bf 9}, 12 (1966) 

\bibitem{Sirovich66} L. Sirovich, {\it Phys. Fluids} {\bf 9}, 2323 (1966) 

\bibitem{ZieringSheinblatt66} S. Ziering \& M. Sheinblatt, {\it Phys. Fluids} {\bf 9}, 1674 (1966) 

\bibitem{GoldmanSirovich67} E. Goldman \& L. Sirovich, {\it Phys. Fluids} {\bf 10}, 1928 (1967) 

\bibitem{LuoGirimaji02} L.-S. Luo \& S. S. Girimaji, {\it Phys. Rev. E}  {\bf 66}, 035301(R) (2001)

\bibitem{LuoGirimaji03} L.-S. Luo \& S. S. Girimaji, {\it Phys. Rev. E} {\bf 67}, 036302 (2003)

\bibitem{GrossJackson59} E.P. Gross \& E.A. Jackson, {\it Phys. Fluids} {\bf 2}, 432 (1959) 

\bibitem{SD_95} X. Shan \& G. Doolen, {\it Jour. Stat. Phys.} {\bf 81}, 379 (1995)

\bibitem{SD_96} X. Shan \& G. Doolen, {\it Phys. Rev. E} {\bf 54}, 3614 (1996)

\bibitem{Shan07} X. Shan, {\it Phys. Rev. E} {\bf 73}, 047701 (2007)

\bibitem{Sbragaglia07}  M. Sbragaglia, R. Benzi, L. Biferale, S. Succi, K. Sugiyama \& F. Toschi, {\it Phys. Rev. E} {\bf 75}, 026702 (2007)

\bibitem{seth91} J. D. Shore, \& J. P. Sethna, {\it Phys. Rev. B} {\bf 43}, 3782 (1991)

\bibitem{seth92} J. D. Shore, M. Holzer \& J. P. Sethna, {\it Phys. Rev. B} {\bf 46}, 11376 (1992)

\bibitem{colloids05} A.I. Campbell, V.J. Anderson, J.S. van Duijneveldt \& P. Bartlett, {\it Phys. Rev. Lett.}  {\bf 94}, 208301 (2005)

\bibitem{Sciortino04} F. Sciortino, S. Mossa, E. Zaccarelli \& P. Tartaglia, {\it Phys. Rev. Lett.} {\bf 93}, 055701 (2004) 

\bibitem{SHAN06} X. Shan, X. F. Yuan \& H. Chen, {\it Jour. Fluid Mech.} {\bf 550}, 413-441 (2006)

\bibitem{Shan08} X. Shan, {\it Phys. Rev. E} {\bf 77}, 066702 (2008) 

\bibitem{NOI09} M. Sbragaglia, R. Benzi, L. Biferale, H. Chen, X. Shan \& S. Succi, {\it Jour. Fluid Mech.} {\bf 628}, 299 (2009)

\bibitem{Sbragaglia09} M. Sbragaglia, H. Chen, X. Shan \& S. Succi, {\it Europhys. Lett.} {\bf 26}, 24005  (2009)

\bibitem{Seul95}  M. Seul \& D. Andelman, {\it Science} 267, 476 (1995).

\bibitem{Stojkovic99} B.P. Stojkovic {\it et al.}, {\it Phys. Rev. Lett.}, {\bf 82}, 4679, (1999)

\bibitem{Reichhardt03} C. Reichhardt, C.J. Olson  Reichhardt, I. Martin \& A.R. Bishop, {\it Phys. Rev. Lett.} {\bf 90}, 026401, (2003)

\bibitem{Tuzel07} E. Tuzel, G. Pan, T. Ihle \& D.M. Kroll, {\it Europhys. Lett.} {\bf 80}, 40010  (2007)

\bibitem{CAV08} G. Biroli, J.-P. Bouchaud, A. Cavagna, T. S. Grigera \& P. Verrocchio, {\it Nat. Phys} 4, 771 - 775 (2008), {\it arXiv:0805.4427v1} 

\bibitem{Coussot02} P. Coussot, J. S. Raynaud, F. Bertrand, P. Moucheront, J. P. Guilbaud, H. T. Huynh, S. Jarny \& D. Lesueur, {\it Phys. Rev. Lett.} {\bf 88}, 218301 (2002)  


\bibitem{Barrat02}  L. Berthier, J-L. Barrat, {\it Phys. Rev. Lett.} {\bf 89}, 095702 (2002) 

\bibitem{BARK} B. Tadjic, {\it Phys. Rev. Lett.}  {\bf 77}, 3843 (1996), and references therein

\bibitem{Becu06} L. Becu, S. Manneville \& A. Colin, {\it Phys. Rev. Lett.} {\bf 96}, 138302 (2006)

\bibitem{Debregeas01} G. Debregeas, H. Tabuteau \& J. di Meglio, {\it Phys. Rev. Lett.} {\bf 87}, 178305 (2001)

\bibitem{Kabla03} A. Kabla \& G. Debr´egeas, {\it  Phys. Rev. Lett.} {\bf 90}, 258303 (2003)

\bibitem{Janiaud06} E. Janiaud, D. Weaire \& S. Hutzler, {\it Phys. Rev. Lett.} {\bf 97}, 38302 (2006)

\bibitem{Salmon03} J. Salmon, A. Colin, S. Manneville \& F. Molino, {\it Phys. Rev. Lett.} {\bf 90}, 228303 (2003)

\bibitem{Lerouge06} Lerouge S., Fardin M.-A., Argentina M., Grégoire G., cardoso O. {\it Soft Matter} {\bf 4}, 1808 (2008); Lerouge S., Argentina M. \& Decruppe J.-P., {\it Phys. Rev. Lett.} {\bf 96}, 088301 (2006)

\bibitem{Huang05} N. Huang, G. Ovarlez, F. Bertrand, S. Rodts, P. Coussot \& D. Bonn, {\it Phys. Rev. Lett.} {\bf 94}, 28301 (2005)

\bibitem{Mills08} P. Mills, P. Rognon \& F. Chevoir, {\it Europhys. Lett.} {\bf 81}, 64005 (2008)

\bibitem{Marze05} S.P.L. Marze, A. Saint-Jalmes \& D. Langevin, {\it Colloids and Surfaces A: Physicochemical and Engineering Aspects} {\bf  263}, 121 (2005) 

\bibitem{BOGOS} M. Nekovee, P.V. Coveney, H. Chen \& B.M. Boghosian, {\it Phys. Rev. E} {\bf 62},  8282-8294 (2000)

\bibitem{Bernaschi09} M. Bernaschi {\it et al.}, in preparation (2009)
























\end{thebibliography}
\end{document}